\documentclass[prd,aps,showpacs,nofootinbib,eqsecnum,amsfonts,amsmath,superscriptaddress,preprintnumbers,floatfix,a4paper,twocolumn]{revtex4}

\usepackage{epsfig}
\usepackage{graphics}
\usepackage{graphicx}
\usepackage{bm}
\usepackage{color}
\usepackage{hyperref}
\usepackage{amssymb}
\usepackage{amsmath}
\usepackage{longtable}
\usepackage{rotating}
\usepackage{color}
\usepackage{epsf}
\usepackage{fancyhdr}

\definecolor{purple}{rgb}{0.5,0,0.5}

\def\bea{\begin{eqnarray}}
\def\eea{\end{eqnarray}}

\newcommand{\be}{\begin{equation}}
\newcommand{\ee}{\end{equation}}
\newcommand{\ber}{\begin{eqnarray}}
\newcommand{\eer}{\end{eqnarray}}
\newcommand{\Mc}{{\cal M}}
\newcommand{\cm}{{\cal M}}
\def\cross{\times}
\def\º{\textrm{\textordmasculine}}

\newcommand{\Ms}{M_{\odot}}

\newcommand{\dt}{{\rm d}t}
\newcommand{\df}{{\rm d}f}
\newcommand{\dN}{{\rm d}N}
\def\oo{\textrm{\textordmasculine}}

\def\inst{\textrm{\mbox{\tiny{inst}}}}
\def\conf{\textrm{\mbox{\tiny{conf}}}}
\def\tot{\textrm{\mbox{\tiny{tot}}}}
\def\eff{\textrm{\mbox{\tiny{eff}}}}
\def\GWD{\textrm{\mbox{\tiny{GWD}}}}
\def\EWD{\textrm{\mbox{\tiny{EWD}}}}
\def\FWF{\textrm{\mbox{\tiny{FWF}}}}
\def\RWF{\textrm{\mbox{\tiny{RWF}}}}

\begin{document}

\title{LISA observations of supermassive black holes: parameter estimation 
using full post-Newtonian inspiral waveforms}


\author{Miquel Trias}
\affiliation{Departament de F\'{\i}sica, Universitat de les Illes
Balears, Cra. Valldemossa Km. 7.5, E-07122 Palma de Mallorca, Spain}

\author{Alicia M. Sintes}
\affiliation{Departament de F\'{\i}sica, Universitat de les Illes
Balears, Cra. Valldemossa Km. 7.5, E-07122 Palma de Mallorca, Spain}
\affiliation{Max-Planck-Institut f\"ur Gravitationsphysik (Albert-Einstein-Institut),
Am~M\"uhlenberg 1, 14476~Golm, Germany}

\date{\today}

\begin{abstract}
We study parameter estimation of supermassive black hole binary systems
in the final stage of inspiral using the full post-Newtonian gravitational waveforms.
We restrict our analysis to systems in circular orbit with negligible spins,
in the mass range  $10^8\Ms-10^5\Ms$, and compare 
the results
with those arising from  the commonly used restricted post-Newtonian approximation. 
The conclusions of this work are particularly important with regard to the astrophysical reach of
future LISA measurements. Our analysis clearly shows that modeling the inspiral with the full
post-Newtonian waveform, not only extends the reach to 
higher mass systems, but also improves in general the 
 parameter estimation. In particular, there are remarkable improvements in angular resolution and distance 
measurement for systems with a total mass higher than $5\times10^6\Ms$, as well as a large 
improvement in the mass determination.
\end{abstract}


\maketitle

\section{Introduction }
\label{sec:intro}
 
Supermassive black hole binary systems,  in the mass range
 $10^8 \Ms - 10^5 \Ms$, will be detectable by the
Laser Interferometer Space Antenna (LISA) \cite{Pre-Phase A Report,Danzmann:2003, lisa}
throughout the entire Universe. Observations of gravitational waves
from this class of sources are among its highest priority targets. 
By measuring these
gravitational waves we will have detailed information regarding 
general relativity itself and the behavior of space-time 
\cite{Dreyer:2003bv,Miller:2004va,Hughes:2004vw,Berti:2004bd,Arun:2006yw,Arun:2006hn};
precision measurements of the Universe as a whole 
\cite{Schutz:1986gp, Hughes:2001ya, Holz:2005df,Kocsis}; 
the formation and  growth of massive black holes in galaxy evolution 
\cite{Hughes:2001ya,Barnes:1992rm,Menou:2001hb}; 
and black hole astrophysics
\cite{Milosavljevic:2004cg,Kocsis:2005vv}.
Detection of gravitational waves will provide complementary information 
to conventional astronomy.

Supermassive black hole binaries are long lived sources in the LISA band.
The whole coalescence of a compact binary system is commonly divided into 
three phases: the adiabatic inspiral, the merger and the ringdown.
Most of the  signal-to-noise ratio (SNR) accumulates during the last 
days prior to coalescence and during the merger phase, but one 
critically relies on long integration 
times to disentangle the source parameters, in particular to resolve the
 source position in the sky
and measure its luminosity distance \cite{Cutler:1997ta}.
This is due to the motion of LISA around the Sun that breaks the
degeneracy in the parameters. Thus, in 
this paper we restrict our attention to the inspiral phase since this is the most 
interesting  for parameter estimation.

Because of the complexity of the problem, most analysis 
carried out so far to address how accurately LISA can measure the source
parameters and the implications for astronomy and cosmology,
have considered only the inspiral restricted post-Newtonian waveforms
 \cite{Cutler:1997ta,Berti:2004bd,Hughes:2001ya,Holz:2005df,Kocsis, Vecchio:2003tn,
 Arun:2006if,Barack:2003fp, Lang:2006bz,Seto:2002uj}, where all amplitude
 corrections are discarded and only post-Newtonian (PN) contributions to the
 phase are taken into account. Thus, the restricted-PN waveform consists of just
 the dominant harmonic at twice the orbital phase. Only in a few cases, it
 has been reported the importance of including higher order PN terms to the
 amplitude and the phase 
 \cite{Sintes:1999ch,Moore:1999zw,Arun:2007qv,Arun:2007p}.
 In the context of ground based detectors, it was found that the consequences
 of amplitude correction in the templates are considerable
 \cite{Sintes:1999cg,Van Den Broeck:2006qu, Van Den Broeck:2006ar}.
 
The main purpose of the present analysis is to investigate in detail 
 the impact of full versus restricted PN waveforms  for parameter
 estimation, by exploring a vast parameter space, 
 specifically in the context of LISA, extending previous results. 
 The waveform we use is described by 11 parameters. Therefore  extensive
 large-scale, CPU-intensive Monte Carlo simulations have been required
 for such an exhaustive study.

The conclusions of this work are particularly important with regard to the astrophysical reach of
future LISA measurements. Our analysis clearly shows that modeling the inspiral with the full
post-Newtonian waveforms, as compared to the restricted-PN ones, not only extends the reach to 
higher mass systems,  up to $10^8\Ms$, 
as discussed by Arun {\it et al.} \cite{Arun:2007qv}, but also improves the 
 parameter estimation. Improvements in angular resolution and distance 
measurement are remarkable for systems with a total mass higher than $5\times10^6\Ms$.
 These results are in agreement with those recently found in \cite{Arun:2007p}.
 
The organization of this paper is as follows. In Section~\ref{sec:noise}, we
discuss the LISA detector output, and we provide the total LISA noise curve 
employed in our analysis. 
In Sec.~\ref{sec:gw} we describe the gravitational wave signals from binary
systems. Sec.~\ref{sec:sape} reviews  the basic concepts of 
signal parameter estimation in matched filtering. 
In Sec.~\ref{sec:bhbs} we spell out the assumptions about the waveform and 
the observations on which our analysis is based, we present a detailed
description of the waveform model and all key steps to compute the SNR and the
estimation of the parameter errors.
Sec.~\ref{sec:results} presents the results of our investigations, where we
compare  the impact on  parameter estimation 
 of supermassive black hole binaries
using full post-Newtonian inspiral waveforms versus the restricted ones.
The results are presented for a given source location, 
exploring a vast parameter space, we also study how source location errors
in terms of advanced warning times, and the errors dependency with redshift.
Finally, Sec.~\ref{sec:summary} concludes with a summary of 
the main results of this paper
and present pointers to future work.


\section{LISA response to gravitational waves and detector noise}
\label{sec:noise}

LISA consists of three spacecrafts arranged in an equilateral triangle orbiting the Sun. 
The arms of the triangle are approximately $L= 5\times 10^6$ km in length, and the triangle is inclined at
an angle of 60$^o$ to the ecliptic. The entire triangular configuration spins as
the antenna orbits the Sun,
rotating once during a single orbit. A gravitational wave interacting with the configuration causes the
length of the three arms to oscillate.

For the supermassive black hole binary inspirals considered in this paper, most of the  SNR accumulates
at frequencies $f<10$~mHz, so it is adequate to use the low-frequency approximation to the LISA response
function derived by Cutler~\cite{Cutler:1997ta}. In this approximation, LISA 
can be regarded as two
independent gravitational wave detectors with 90$^o$ arms and rotated 45$^o$ with respect to one another.

The strain $h(t)$ produced at the output of LISA Michelson interferometer  by a gravitational wave (GW) signal
characterized by two polarization states $h_+(t)$ and $h_\times(t)$ is
\begin{equation}
h^{(i)}(t) = {\sqrt{3} \over 2}\left[F_+^{(i)}(t)h_+(t) + F_\times^{(i)}(t)h_\times(t) \right]
\, ,
\end{equation}
where $F_+^{(i)}$ and $F_\times^{(i)}$ are the time-dependent  antenna 
pattern functions, the factor ${\sqrt{3} / 2}$ comes from the 60$^o$ opening angle of the LISA arms, and
the $i=$I,II labels the two independent Michelson outputs.
The response functions $F_+^{(i)}(t)$ and $F_\times^{(i)}(t)$ depend on the 
direction and orientation of the source in the sky and they vary with time because during the observation,
the interferometer changes orientation with respect to the source. 
We refer the reader to \cite{Cutler:1997ta} for further discussions and details.

 The total noise that affects the  observation of radiation emitted by binary systems
 is given by the superposition of instrumental sources, $S_n^\inst(f)$, and astrophysical
 foregrounds of unresolved radiation, $S_n^\conf(f)$, the so-called confusion noise.
The total noise spectral density $S_n(f)$ is therefore the sum of these two components
\be
S_n(f)=  S_n^\inst(f)+ S_n^\conf(f) \, .
\ee
The noise contributions in each data stream I and II are by definition the same.

A good fit for the instrumental noise, for $f\le 5$~mHz, is given by~\cite{Barack:2004wc,Finn:2000sy}
\bea
S_n^\inst(f) &= &6.12\times 10^{-51}~f^{-4} + 1.06\times 10^{-40} \nonumber\\
& & +6.12\times 10^{-37}~f^2~ \mathrm{Hz}^{-1}
\eea
where $f$ is in Hz. This is derived from the \textit{online sensitivity curve generator}~\cite{larson},
which is based on the noise budgets specified in~\cite{Pre-Phase A Report}.

For the confusion noise we adopt the same analytical approximations given
 in~\cite{Barack:2004wc},
considering only noise from short-period galactic $S_n^{\GWD}$ and extragalactic 
binaries $S_n^{\EWD}$ (due to white dwarfs binaries), assuming they are all unresolvable, i.e.:
the worst case,
and we ignore the effects
of captures of compact objects. That is, we estimate the total effective noise
density as
\bea
\label{Eq.Effective_noise}
S_n^{\eff}(f) &=& \min\left\lbrace
 \left[S_n^{\inst}(f) + S_n^{\EWD}(f)\right]\, \exp(\kappa T^{-1} \dN/\df),
 \right. \nonumber\\
  & & \left. S_n^{\inst}(f) + S_n^{\EWD}(f) + S_n^{\GWD}(f) \right\rbrace
\eea
where we take $\kappa\,T^{-1} = 1.5$~yr$^{-1}$,
\be
\dfrac{\dN}{\df} = 2\times 10^{-3}~f^{-11/3}~ \mathrm{Hz}^{-1}\, ,
\ee
\begin{equation}
S_n^{\GWD}(f) = 1.4\times 10^{-44} ~f^{-7/3}~ \mathrm{Hz}^{-1}\, , 
\end{equation}
and
\begin{equation}
S_n^{\EWD}(f) = 2.8\times 10^{-46} ~f^{-7/3}~\mathrm{Hz}^{-1}\, .
\end{equation}


\section{Binary black hole coalescence waveforms}
\label{sec:gw}

The coalescence of binary black holes is commonly divided 
into three successive epochs in the
time domain: inspiral, merger and ringdown. During the inspiral the distance
between the black holes diminishes and the orbital frequency sweeps up.
The waveforms are well modeled using the post-Newtonian
approximation to general relativity.
Eventually the post-Newtonian
description of the orbit breaks down, and the black holes cannot be
treated as point particles any more. What is more, it is expected that they
will reach the \emph{innermost stable circular orbit} (ISCO), at which the
gradual inspiral ends and the black holes begin to plunge together to
 form a single black hole.
This is referred as the merger phase. At the end,
 the final black hole will gradually settle down into a Kerr black hole.
 
 In recent years, a series of breakthroughs has occurred in numerical
 simulations of binary black hole systems 
 \cite{Pretorius:2005gq,Campanelli:2005dd,Baker:2005vv}. Long-term 
 evolutions  of inspiralling black holes that last for several orbits have been
 obtained with several independent codes
 \cite{Pretorius:2006tp,Campanelli:2006gf,Baker:2006yw,Sperhake:2006cy, 
 Bruegmann:2006at, Scheel:2006gg, Herrmann:2007ac,
 Koppitz:2007ev,Husa:2007rh,Husa:2007hp},  and accurate GW signals have been
 computed, including the merger and ringdown phases. Still, the post-Newtonian
 approximation to general relativity 
 is the best available method for calculating the vast majority
 of the GW signal cycles observed by LISA. See \cite{Blanchet:2002av} for a review and 
 extensive references.
 
 For non-spinning black holes, the best PN waveforms currently available
 have been calculated at 2.5PN order in amplitude and 3.5PN order in phase
 \cite{Blanchet:1995fr,Blanchet:1995ez,Blanchet:1996pi,Blanchet:2001ax,
 Blanchet:2004ek,Nissanke:2004er}. These waveforms in the two polarizations 
  $h_+$ and $h_{\times}$  take the general form
\bea
\label{expansion}
h_{+,\times}&=& {2M\eta\over D_L}(M\omega)^{2/3} \left\{H_{+,\times}^{(0)}
+v^{1/2}H_{+,\times}^{(1/2)}+v H_{+,\times}^{(1)} \right.\nonumber\\
& & \left.
 +v^{3/2}H_{+,\times}^{(3/2)}
+v^2 H_{+,\times}^{(2)}+ v^{5/2}H_{+,\times}^{(5/2)} \right\} \ ,
\eea
where we have set  $G=c=1$, as we will do throughout this paper,
 $v\equiv (M\omega)^{2/3}$, $\omega$ is the  orbital frequency,
$D_L$ is the luminosity distance to the source, and $M$ and $\eta$ are the observed total mass
and the symmetric mass ratio respectively, defined in 
Sec.~\ref{sec:bhbs}.
The explicit expressions for $H_{+,\times}^{(m/2)}$, $m=0,\ldots,5$ can be found in
\cite{Blanchet:1996pi,Arun:2004ff}. They include contributions from several harmonics
of the binary's orbital motion.

For black holes with significant spins, the state of the art is somehow less advanced and 
the corresponding waveforms have been calculated through 2.5PN order 
\cite{Faye:2006gx,Blanchet:2006gy}.

 Equation~(\ref{expansion}) corresponds to the 
so-called \textit{full} waveform (FWF). Given its complexity, together with the
 fact that the second harmonic contributes most strongly to the waveform over most
 of the inspiral phase, it is common to make some simplifications and work only 
 with the \textit{restricted} waveform (RWF), in which one neglects all 
 amplitude terms
 except the Newtonian quadrupole one, but keeping the phase to some specific PN
 order, i.e., keeping only $H_{+,\times}^{(0)}$ and throwing out the rest
  $H_{+,\times}^{(m/2)}$ for  $m>0$.

It is the goal of this paper to revisit the problem of parameter estimation 
for supermassive black hole binaries studying the improvement in error estimation by
using FWF and compare with the previous results obtained with the RWF.

\section{Review of signal analysis and parameter estimation}
\label{sec:sape}

In this section we briefly review the basic concepts and formulas of
signal  parameter estimation relevant to the goal of this paper;
we refer the reader to \cite{Cutler:1994ys} for a more detailed analysis.

The signal $s^{(i)}(t)$ as measured by the detector $i$ is a superposition of 
noise $n^{(i)}(t)$ and gravitational waves $h^{(i)}(t;{\bm \lambda})$
\be
s^{(i)}(t)=h^{(i)}(t;{\bm \lambda})+n^{(i)}(t)
\ee
where ${\bm \lambda}$ represents a vector of the unknown parameters (location, masses, spins,
etc) that characterize the actual waveform and that one wishes to estimate from the data
stream. 

For sake of simplicity we shall made the standard
assumptions that the noise $n^{(i)}(t)$ has zero mean and it 
is stationary and Gaussian.
Within this approximation, the Fourier components 
of the noise
are statistically described by
\be
E[\tilde{n}(f)\tilde{n}^*(f')] =  \frac{1}{2} \delta (f-f')S_n(f) \, ,
\ee
where $E[]$ denotes the expectation value with respect to an ensemble of noise
realization, the $*$ superscript denotes complex conjugate,
$S_n(f)$ is the one sided  noise power spectral density, and 
tildes denote Fourier transforms according to the convention
\be
\label{fourier_transform}
\tilde{x}(f) = \int_{-\infty}^{\infty}  e^{i2 \pi f t} x(t)\;  \dt \, .
\ee
%

With a given noise spectral density for the detector, one defines
the ``inner product'' between any 
two signals $g(t)$ and $h(t)$ by
\begin{equation}\label{inner_product}
( g | h ) \equiv 2 \int_0^{\infty} \frac{\tilde{g}^*(f) \tilde{h}(f)
+ \tilde{g}(f) \tilde{h}^*(f)}{S_n (f)} \; \df \, .
\end{equation}
With this definition, the probability of the noise to have a
realization $n_0$ is just
\be
p(n=n_0) \propto e^{-(n_0 | n_0)/2} \, .
\ee

The optimal signal-to-noise ratio (SNR) $\rho$, achievable with linear methods
({\it e.g.}, matched filtering the data) is given by the standard expression
\begin{equation}
\label{snr_2}
\rho^2=( h | h ) = 4 \int_0^{\infty} \frac{|\tilde{h}(f)|^2}{S_h(f)}\; \df \, .
\end{equation}

In the limit of large SNR, which is typically the case for LISA observations of 
supermassive black
hole binary systems,
 the probability that the gravitational wave signal $h(t;{\bm \lambda})$ is characterized by a
given set of values of the source parameters 
${\bm \lambda}=\{\lambda^k\}$ is given by a Gaussian
probability  of the form \cite{Finn:1992wt}
\be
\label{p10}
p( \mbox{\boldmath $\lambda$} \vert h)=p^{(0)}(\mbox{\boldmath $\lambda$})
 \exp \left[ - \frac{1}{2} \Gamma_{j k} \Delta\lambda^j \Delta\lambda^k \right] \,
 \, ,
\ee
where $\Delta\lambda^k$ is the difference between the true value of the
parameter and the best-fit parameter in the presence of some realization of the
noise, 
$p^{(0)}(\mbox{\boldmath $\lambda$})$ represents the distribution of prior
information (a normalization constant) and 
 $\Gamma_{j k}$ is the so-called Fisher information matrix defined by
\bea
\label{fisher_matrix}
\Gamma_{i j} &&\equiv (\partial_i h | \partial_j h) \\
&& = 2 \int_0^{\infty} \frac{\partial_i \tilde{h}^*(f)
\partial_j \tilde{h}(f)
+ \partial_i \tilde{h}(f) \partial_j \tilde{h}^*(f)}{S_n (f)} \; \df \, , \nonumber
\eea
where $\partial_i = \frac{\partial}{\partial \lambda^i}$.

The inverse of the Fisher matrix, known as the 
variance-covariance matrix, gives us the accuracy with 
which we expect to measure
the parameters $\lambda^k$
\begin{equation}
\label{covariance_matrix}
\Sigma^{jk} \equiv 
(\Gamma^{-1})^{jk} = \langle \Delta\lambda^j\Delta\lambda^k\rangle \, .
\end{equation}
Here the angle brackets denote an  average over the probability
distribution function in Eq.~(\ref{p10}).
The root-mean-square error $\sigma_k$ in the estimation of the parameters
$\lambda^k$ can then be calculated, in the limit of large SNR, by taking the
square root of the diagonal elements of the variance-covariance matrix,
\be
\sigma_k = \langle (\Delta\lambda^k)^2\rangle^{1/2} =\sqrt{\Sigma^{kk}} \, ,
\ee
and the correlation coefficients  $c^{jk}$ between two 
parameters $\lambda^j$ and
$\lambda^k$ are given by
\be
c^{jk}= \frac{\langle \Delta\lambda^j\Delta\lambda^k\rangle}{\sigma_j\sigma_k}=
\frac{\Sigma^{jk}}{\sqrt{\Sigma^{jj}\Sigma^{kk}}} \, .
\ee

Returning again to the two detector case, we have that the largest value
of the SNR is
\be
\rho^\tot= \sqrt{(\rho^I)^2+(\rho^{II})^2} \, ,
\ee
\be
(\rho^{I,II})^2 = ( h^{I,II} | h^{I,II} ) \, ,
\ee
and we can write a total Fisher matrix as the sum of the individual Fisher matrices for each
detector
\be
\Gamma_{i j}^\tot=\Gamma_{i j}^I+ \Gamma_{i j}^{II} \, .
\ee

In this paper we will use equation (\ref{covariance_matrix}), along with the  FWF
and RWF models for the binary black hole coalescence, and
the LISA noise spectrum  discussed in Sec.~\ref{sec:noise}, in order to estimate how well 
LISA will be able to measure the source parameters. 
We refer the reader to
\cite{Cutler:2007mi,Vallisneri:2007ev} for a detailed discussion about the conditions required
for the Fisher-matrix formalism to be applicable.


\section{Observation of supermassive black hole binary systems }
\label{sec:bhbs}

We consider observations of supermassive black hole binary systems
of masses $m_1$ and $m_2$ at luminosity distance $D_L$. For later convenience we define the
following mass parameters: total mass $M=m_1+m_2$, reduced mass
$\mu=m_1m_2/M$,  the symmetric mass
ratio $\eta=\mu/M$, and ${\cal M}=\mu^{3/5}M^{2/5}=M\eta^{3/5}$  the chirp mass.
We focus on binary systems in the mass range  $10^8 \Ms - 10^5 \Ms$ and  we do not consider here
the case of binaries with an extreme mass ratio, e.g, a black hole of $10 \Ms$ orbiting a
supermassive one. The reason is that  some assumptions about the waveform that we will be
considering would be rather unrealistic for such astrophysical system.

Before presenting the results, in this section we spell out the assumptions about the waveform
and the observations on which our analysis is based, together with a detailed description of the 
post-Newtonian inspiral waveform we use.

\subsection{Assumptions}
\label{sec:assumptions}

The waveform model we consider is based on the following assumptions:
\begin{itemize}

\item As signal we consider only the inspiral phase of the whole coalescence, neglecting all
information coming from the merger and the ringdown phases. We terminate the inspiral when the
binary's members are separated by a distance $6M$; this very roughly corresponds to the point at
which the post-Newtonian expansion ceases to be accurate.

\item We restrict ourselves to circular orbits by omitting the orbital eccentricity;
 this hypothesis is considered rather realistic for 
 the supermassive binary systems visible in the LISA band we consider.

\item The contributions of spins are negligible or they are oriented in such a way that no
spin-induced precession of the orbital plane takes place; we take care of spin contributions
only into the waveform phase. We choose this hypothesis in order to control the complexity of
the problem and focus on the comparison of the FWF versus RWF.

\item We approximate the waveform at the 2PN order, both in amplitude and in phase, considering
up to six harmonics in the case of FWF. This simplification is motivated by computational
reasons due to our limited computational resources and does not affect in any significant way
the final results.

\item Out of the seventeen parameters on which the most general waveform depends on, the 2PN
approximation we are considering here depends only on eleven parameters: the luminosity
distance, four angles defining the constant source position and orientation of the binary
in the orbital plane, two mass parameters, two parameters related to the spin-orbit and spin-spin
coupling and one arbitrary phase and time.

\end{itemize}

We assume that the observations are carried out according to the following:

\begin{itemize}
\item We consider sources at cosmological distances, as the event rate of massive black hole binary
systems in our local Universe is likely to be negligible and only taking into account the whole Universe
it becomes of significant importance; moreover there is a great interest in carrying high redshift
surveys; indeed, unless differently stated, we consider fiducial sources to be 
at redshift $z=1$, in a flat
Universe described by the following cosmological parameters: 
 $H_0 = 71~\mathrm{km}~\mathrm{s}^{-1}~\mathrm{Mpc}^{-1}$,
 $\Omega_m = 0.27$ and $\Omega_\Lambda = 0.73$; the corresponding luminosity distance is
therefore
\begin{equation} \label{Eq.DL(z)}
D_L(z) = \frac{1+z}{H_0}  \int_0^z\dfrac{dz'}{\sqrt{\Omega_m (1+z')^3 + \Omega_{\Lambda}}} \, .
\end{equation}
All the parameters considered here are the \textit{observed ones}; they differ from the values of
the parameters as measured in the source rest frame according to
\bea
f&\rightarrow& \frac{f}{1+z} \nonumber\\
t &\rightarrow& (1+z)t \nonumber\\
M &\rightarrow& (1+z)M \nonumber\\
\Mc &\rightarrow& (1+z)\Mc \nonumber\\
\mu &\rightarrow& (1+z)\mu\, .
\eea

\item  Unless differently stated, we consider that LISA observes the inspiral for a 
whole year before it reaches the ISCO. This corresponds to different frequency ranges depending on
the harmonic. For every choice of source parameters these frequencies are computed in advanced
and we also impose a low-frequency cut-off to the instrument at $5\times 10^{-5}$~Hz.

\item The total noise that affects the observation is given  by Eq.~(\ref{Eq.Effective_noise}),
that is, we take both the instrumental and confusion noise contributions.

\item We compute the  expected mean square errors $\langle (\Delta\lambda^k)^2\rangle^{1/2}$
and the angular resolution of the instrument, which we define as 
\be
\Delta\Omega_N= 2\pi \sqrt{ \langle \Delta\cos\theta_N^2\rangle 
\langle \Delta\phi_N^2\rangle -\langle \Delta\cos\theta_N \Delta\phi_N\rangle^2} \, ,
\ee
where $(\theta_N, \phi_N)$ are the polar angles in the solar system barycentre frame
of the source location in the sky,
with one and with both detectors, but we
only report the results for the combined case. 

\item The analysis is done in the frequency domain using the stationary phase
approximation:
we first compute analytically the derivatives $\partial_j \tilde{h}^{(i)}$, where $j=1,\ldots,11$,
then compute numerically the Fisher matrix $\Gamma_{i j}$ and  the variance-covariance matrix 
$\Sigma^{jk}$; the integration and matrix inversion are performed using numerical 
routines of the \texttt{gsl} library.

\item To provide an overall picture of the instrument performances, we study not just a few
cases in detailed but we also do extensive Monte-Carlo simulations, by varying the relevant
source parameters, in particular the position and orientation of the source,
as it turns out  to affect very significantly the parameter measurements. 

\end{itemize}

\subsection{The post-Newtonian inspiral waveforms}
\label{sec:pn}

In this section we derive explicit ready-to-use analytical expressions for the
signal measured at the LISA detector output for inspiral binary systems in circular orbit
within the 2PN approximation, that could easily be expanded to include higher order terms.

For our analysis, 
%
%
it is convenient to expand $h^{(i)}(t)$ as a summation of
different multipole terms, which can be written schematically as
\bea \label{Eq.GW_1}
h^{(i)}(t) &=& \sum_{j=1}^6 h^{(i)}_j(t) = 
\sum_{j=1}^6  \dfrac{\sqrt{3}}{2} 2 M\eta \dfrac{1}{D_L}x^2
\nonumber \\
 & & \times\left[ \left( u_{+,j} F_+ + u_{\cross,j} F_\cross \right)
\cos(\frac{j}{2}\Phi + \varphi_D) \right. \nonumber \\
 & & + \left. \left( w_{+,j} F_+ + w_{\cross,j} F_\cross  \right)
\sin(\frac{j}{2}\Phi + \varphi_D)\right] \, ,\nonumber \\
 & &
\eea
where 
$\varphi_D$ is the LISA's
Doppler phase, $\left\lbrace u_{+,\cross}, w_{+,\cross}\right\rbrace$ contain an internal
summation over all PN orders (see Eq.~(\ref{Eq.us_ws}) below), $x$ is the PN expansion parameter
\begin{equation}\label{Eq.xdef}
x \equiv \left(M w \right)^{1/3}  \, ,
\end{equation}
being $w$ the orbital frequency and $\Phi$ is given in Eq.~(\ref{Eq.Phi(F)}).

In general, $u_{(+,\cross),j}$ and $w_{(+,\cross),j}$ can be written (at least, up to 2PN) as
Eq.~(\ref{Eq.us_ws}). Analyzing those terms, given explicitely in \cite{Blanchet:1996pi},
 one realizes that all of them have a common factor $\Upsilon(j)$ that only 
 depends on the multipole $j$ we are working with, regardless of the PN order
 considered
 %
%
\begin{eqnarray} \label{Eq.us_ws}
u_{(+,\cross),j} &=& \sum_{n=0}^4 x^n u_{(+,\cross),j}^{(n)} 
= \Upsilon(j) \sum_{n=0}^4 x^n \hat{u}_{(+,\cross),j}^{(n)} \nonumber \\
 &&\equiv \Upsilon(j) \, \hat{u}_{(+,\cross),j} \nonumber \\
w_{(+,\cross),j} &=& \sum_{n=0}^4 x^n w_{(+,\cross),j}^{(n)} 
= \Upsilon(j) \sum_{n=0}^4 x^n \hat{w}_{(+,\cross),j}^{(n)}  \nonumber \\
&&\equiv \Upsilon(j) \, \hat{w}_{(+,\cross),j}  \, .
\end{eqnarray}
This notation is very convenient for computing analytically the derivatives of the waveform.
All $(\hat{u}, \hat{w})_{(+,\cross),j}^{(n)}$ can be found in
Appendix~\ref{Sec.us_ws}. The  factors $\Upsilon(j)$, for each multipole,
are
\begin{eqnarray} \label{Eq.ExtraFactor}
\Upsilon(j=1) &\equiv& s \dfrac{\delta m}{M} \nonumber\\
\Upsilon(j=2) &\equiv& 1 \nonumber\\
\Upsilon(j=3) &\equiv& s \dfrac{\delta m}{M} \nonumber\\
\Upsilon(j=4) &\equiv& s^2 \nonumber\\
\Upsilon(j=5) &\equiv& s^3 \dfrac{\delta m}{M} \nonumber\\
\Upsilon(j=6) &\equiv& s^4 \, ,
\end{eqnarray}
%
where 
$s\equiv\sin\iota=\mid\hat{\textbf L}\times\hat{\textbf N}\mid$ and 
$c\equiv\cos\iota=-\hat{\textbf L}\cdot\hat{\textbf N}$.
The source location in the sky $\hat{\textbf N}$ and the orbital angular momentum
$\hat{\textbf L}$ can be described  by the polar angles $(\theta_N, \phi_N)$ and $(\theta_L,
\phi_L)$ with respect to the solar system barycentre frame.

%
%

With all these considerations, the 
gravitational waveform, given by Eq.~(\ref{Eq.GW_1}), can be rewritten as follows
\begin{equation}\label{Eq.GW_2}
h^{(i)}_j(t) = \dfrac{\sqrt{3}}{2} 2 M\eta \dfrac{1}{D_L}x^2 A_j \cos\left(\frac{j}{2}\Phi +
\varphi_{p,j} + \varphi_D \right) \, ,
\end{equation}
where
\begin{eqnarray} \label{Eq.A_phi_defs}
A_j &=& \mid \Upsilon(j) \mid \left[ \left( \hat{u}_{+,j} F_{+} + \hat{u}_{\cross,j}
F_{\cross}\right)^2 \right. \nonumber \\
 & & + \left. \left( \hat{w}_{\cross,j} F_{\cross} + \hat{w}_{+,j}
F_{+}\right)^2\right]^{1/2} \, ,
\nonumber \\
\varphi_{p,j} &=& \tan^{-1} \left[ \dfrac{-\left(\hat{w}_{\cross,j} F_{\cross} + \hat{w}_{+,j}
F_{+}\right)}{\left( \hat{u}_{+,j} F_{+} + \hat{u}_{\cross,j} F_{\cross}\right)} \right] \, .
\end{eqnarray}
Note that $\Upsilon(j)$ cancels in $\varphi_{p,j}$ expression, which prevents a lot of
divergence problems in the numerical computation of the Fisher matrix. 
See  Appendix~\ref{Sec.tfp_evolution} for further details regarding
 the waveform in the frequency
domain.

%
%
%


\section{Results }
\label{sec:results}

The signal model considered here depends on eleven independent parameters, and
it is rather natural to consider the choice
\be
{\bm \lambda}=\{ \cos \theta_N, \phi_N ,\cos \theta_L,\phi_L,\ln D_L, 
 t_c, \phi_c, \beta,
\sigma, \ln \Mc, \ln \mu \}
\ee
that we adopt in this study, in order to easy comparison with previous existing
results.  $t_c$ and $\phi_t$ are the time and phase at coalescence, and $\beta$ 
and $\sigma$ are the so called spin-orbit and spin-spin parameters.
We usually set $t_c=\phi_c=\beta=\sigma=0$ in our analysis.

The code we use  to compute the Fisher matrix is
an extension of a previous one developed by Alberto Vecchio. Our code allows for multiple 
choices: selection of PN order in amplitude and phase, 
number of harmonics, and also different sets of independent parameters. 
In particular, for the FWF it is more convenient to consider
a different choice of the mass parameters to avoid divergences. Therefore, 
sometimes we consider the mass parameter combination 
$\{ \delta m, M\}$, where $\delta m= m_1-m_2$ instead of $\{\Mc,\mu\}$, although
we always express the results for comparison with respect to the later choice. The
relation of the variance-covariance matrix components, with respect
to the different parameters, are
\bea
\label{6.2}
\Sigma^{\Mc\Mc} &=& 
\left( \frac{\partial \Mc}{\partial \delta m}\right)^2\Sigma^{\delta m\delta m} +
\left( \frac{\partial \Mc}{\partial M}\right)^2\Sigma^{MM} \nonumber\\
& +& 2 \frac{\partial \Mc}{\partial M}\frac{\partial \Mc}{\partial \delta m}\Sigma^{M\delta m} \\
\Sigma^{\mu\mu} &=& 
\left( \frac{\partial \mu}{\partial \delta m}\right)^2\Sigma^{\delta m\delta m} +
\left( \frac{\partial \mu}{\partial M}\right)^2\Sigma^{MM} \nonumber\\
& +&2  \frac{\partial \mu}{\partial M}\frac{\partial \mu}{\partial \delta m}\Sigma^{M\delta m} \\
\Sigma^{\Mc\mu} &=& 
\frac{\partial \Mc}{\partial \delta m} \frac{\partial \mu}{\partial \delta m}
\Sigma^{\delta m\delta m} +
\frac{\partial \Mc}{\partial M} \frac{\partial \mu}{\partial M}
\Sigma^{MM} \nonumber\\
&+&  \left( \frac{\partial \Mc}{\partial M}\frac{\partial \mu}{\partial \delta m}
+  \frac{\partial \Mc}{\partial \delta m} \frac{\partial \mu}{\partial M}\right)
\Sigma^{M\delta m} 
\eea
where
\be
\frac{\partial \Mc}{\partial M}= \frac{\Mc}{20\mu}\left[5 +\frac{\delta m^2}{M^2} \right]
\, ,
\ee
\be
\frac{\partial \Mc}{\partial \delta m}= -\frac{3\Mc}{10\mu}\frac{\delta m}{M} \, ,
\ee
\be
\dfrac{\partial\mu}{\partial M} = \dfrac{1}{4} + \dfrac{\delta m^2}{4 M^2} \, ,
\ee
\be
\label{6.8}
\dfrac{\partial\mu}{\partial\delta m} = -\dfrac{\delta m}{2 M} \, .
\ee

Unfortunately, for the equal mass case, the Jacobian of the transformation between
$\{ \delta m, M\}$ and  $\{\Mc,\mu\}$ is singular, and also the
Fisher matrix presents a coordinate singularity  depending on the choice of mass parameters
and the waveform model used. Since we are still interested in the errors in $\ln \Mc$ and $\ln\mu$
we convert them using the same previous formulas. For the unequal masses, we find that computing the
errors in  $\{ \delta m, M\}$ and then converting gives the same result as simply computing the
errors in  $\{\Mc,\mu\}$ directly. Further details are discussed in 
Appendix~\ref{sec:equalmass}.

We have checked that our results agree with Vecchio's code for the RWF, and
with Sintes and Vecchio~\cite{Sintes:1999ch} at the  0.5PN-2PN order in 
amplitude and phase, respectively. We use numerical integration and matrix inversion
routines from the \texttt{gsl} library that, for some particular cases, we have checked against
 \texttt{Mathematica}.
 
 We work with the full $11\times 11$ Fisher matrix for both RWF and FWF.
 There are parameter configurations for which either the numerical integrations or
 the inversion of such a large matrix tend to
 fail (or the results do not have the desired accuracy). The reason is that the Fisher matrix
 is often ill conditioned.  Because of that, we have limited our study up to a mass ratio
 $m_2/m_1=0.01$.

\subsection{The impact of the FWF: General trends}

\begin{figure}
\begin{center}
\includegraphics[width=9cm]{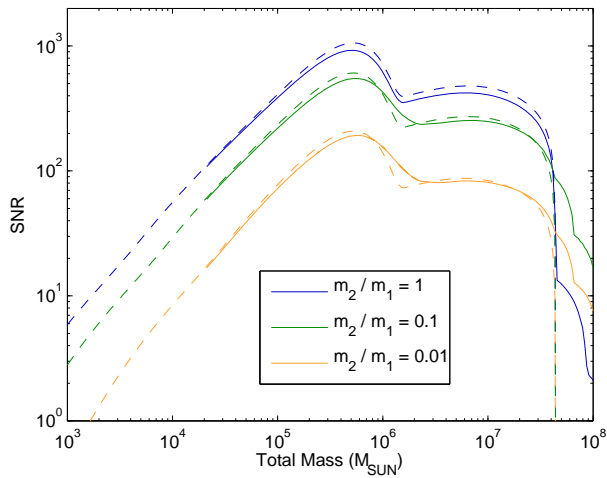}
\end{center}
\caption{SNR versus total mass for the mass ratios of 1, 0.1 and 0.01 for an integration time of
one year.
The solid lines correspond the FWF and dashed lines to RWF. 
The sources are at redshift $z=1$, corresponding to a luminosity
distance of $D_L= 6.64$~Gpc, with fixed angles given by  
$\cos \theta_N = -0.6$, $\phi_N = 1$, $\cos \theta_L = 0.2$ and  
$\phi_L = 3$.}
\label{Fig.FixedAng_SNR}
\end{figure}

\begin{figure*}
\begin{tabular}{cc}
\includegraphics[width=7.5cm]{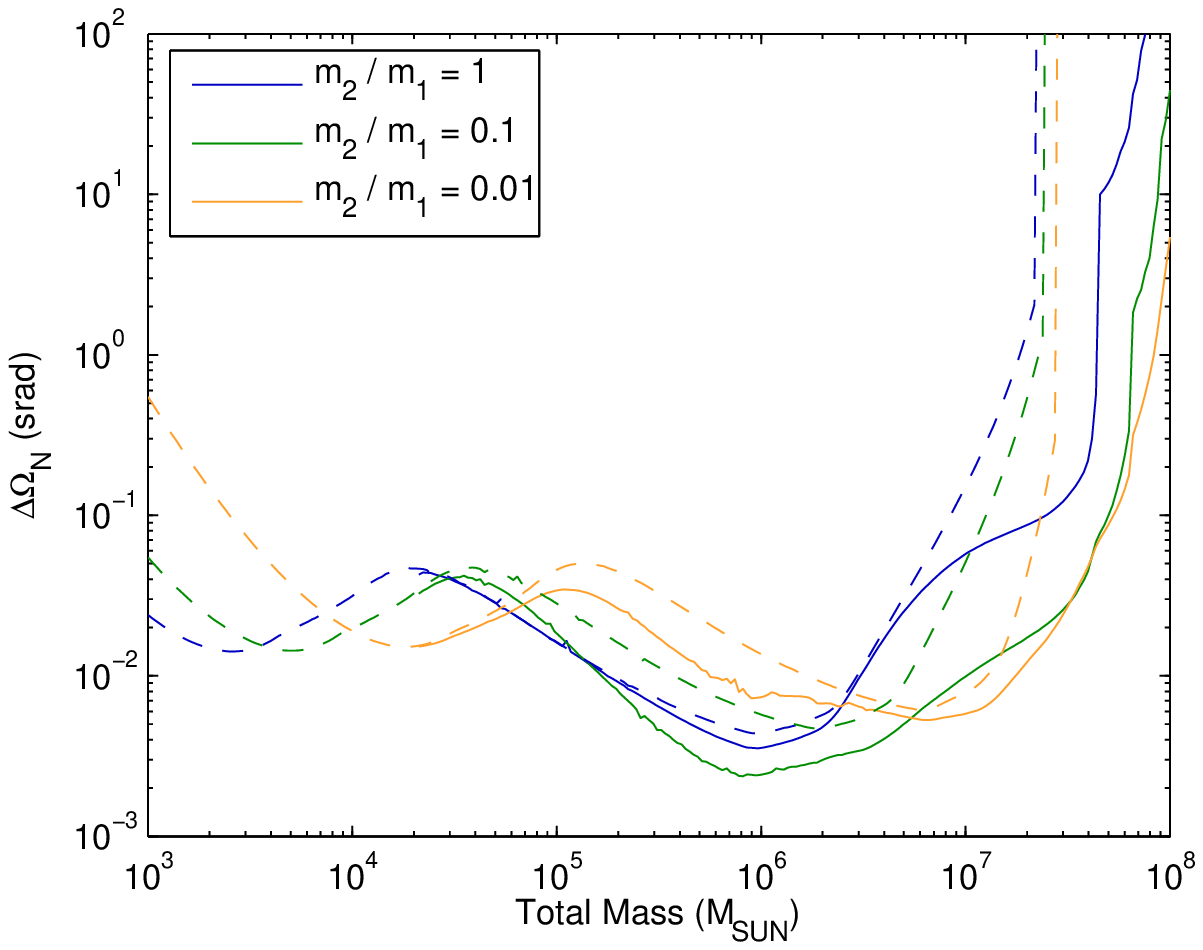} & \includegraphics[width=7.5cm]{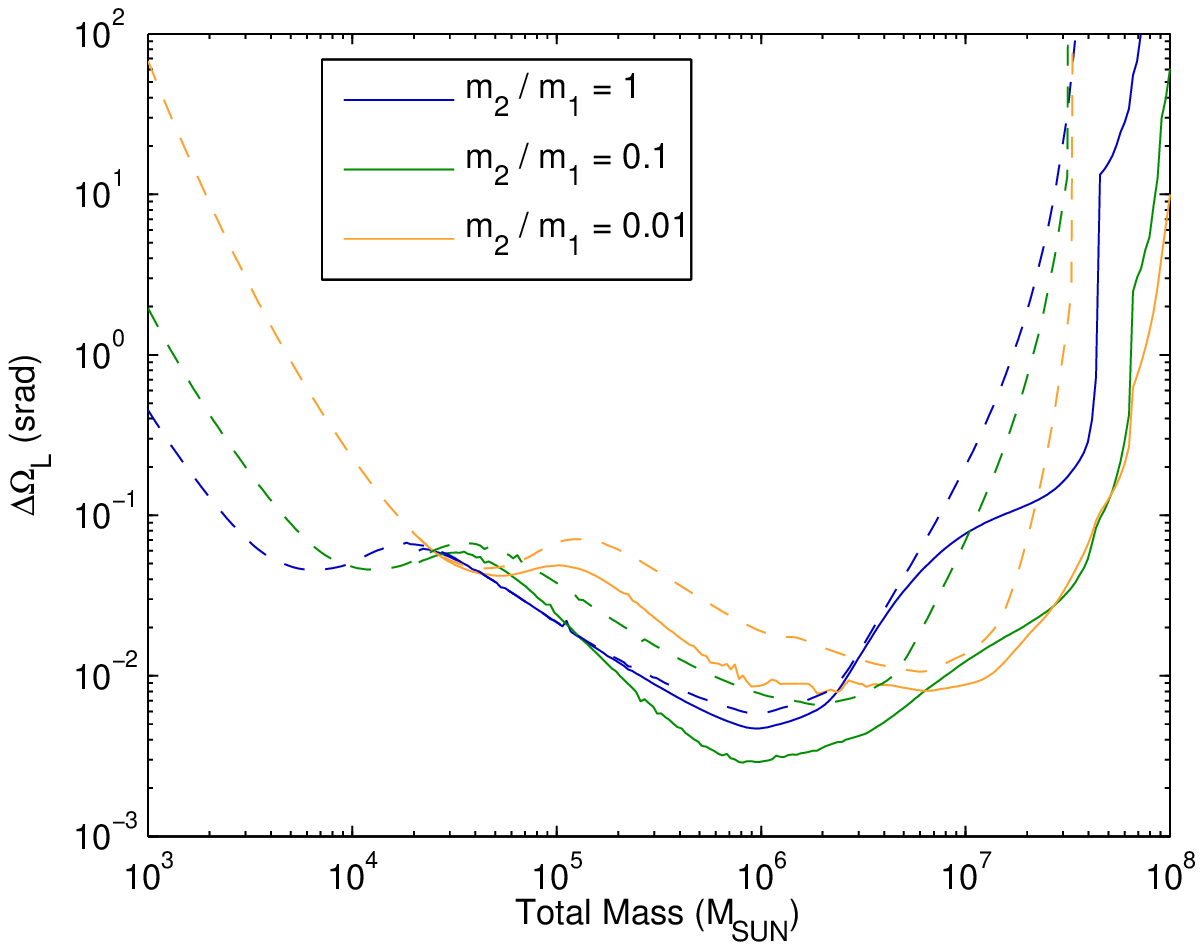} \\
\includegraphics[width=7.5cm]{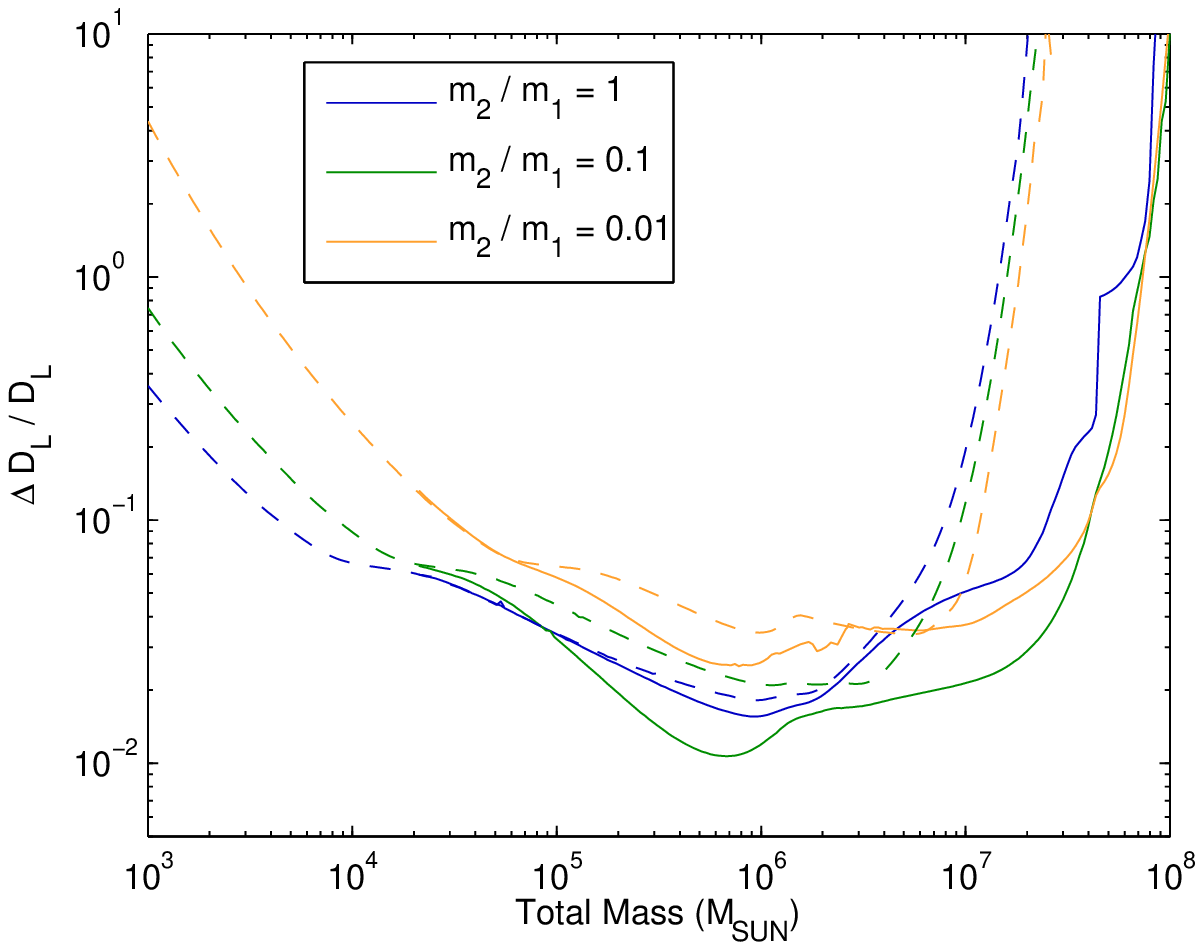} & \includegraphics[width=7.5cm]{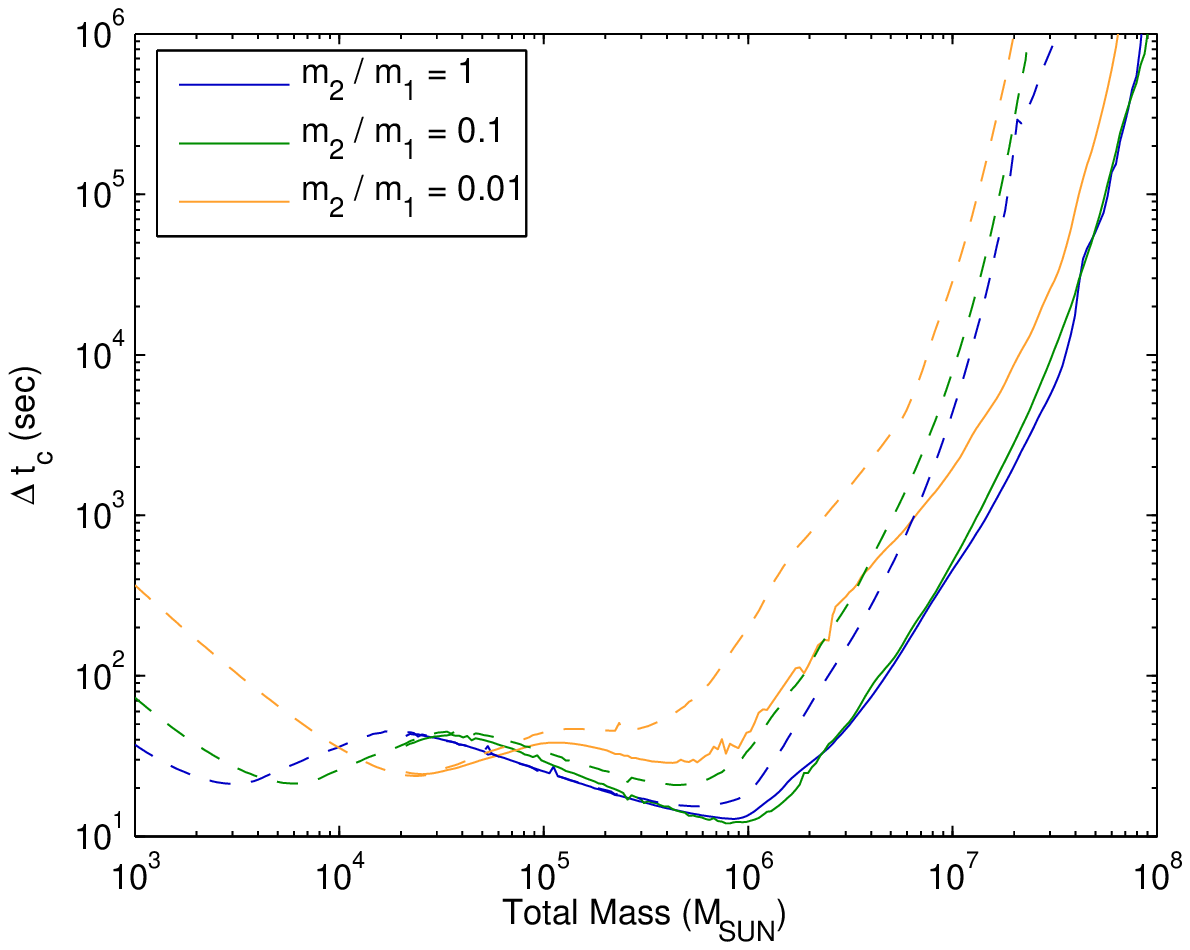} \\
\includegraphics[width=7.5cm]{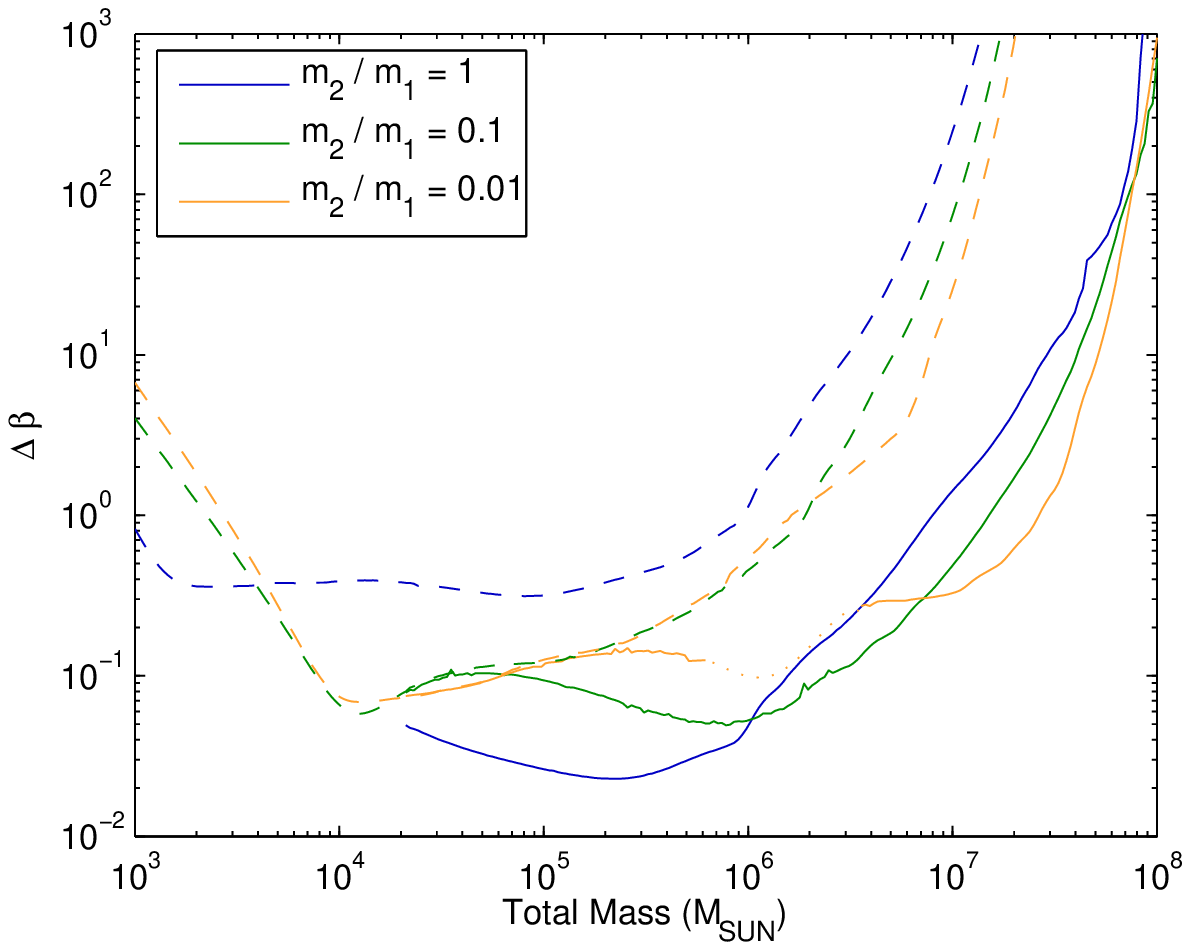} & \includegraphics[width=7.5cm]{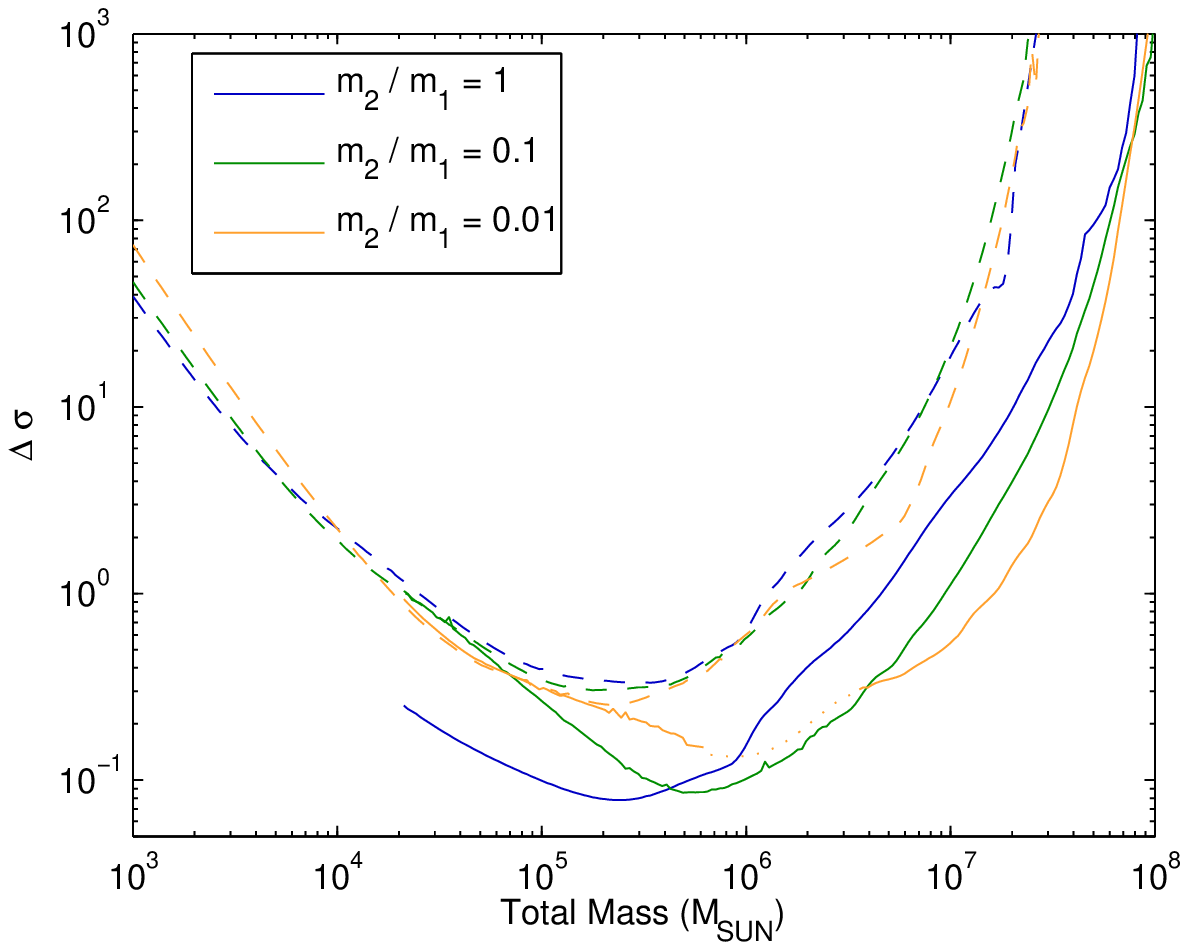} \\
\includegraphics[width=7.5cm]{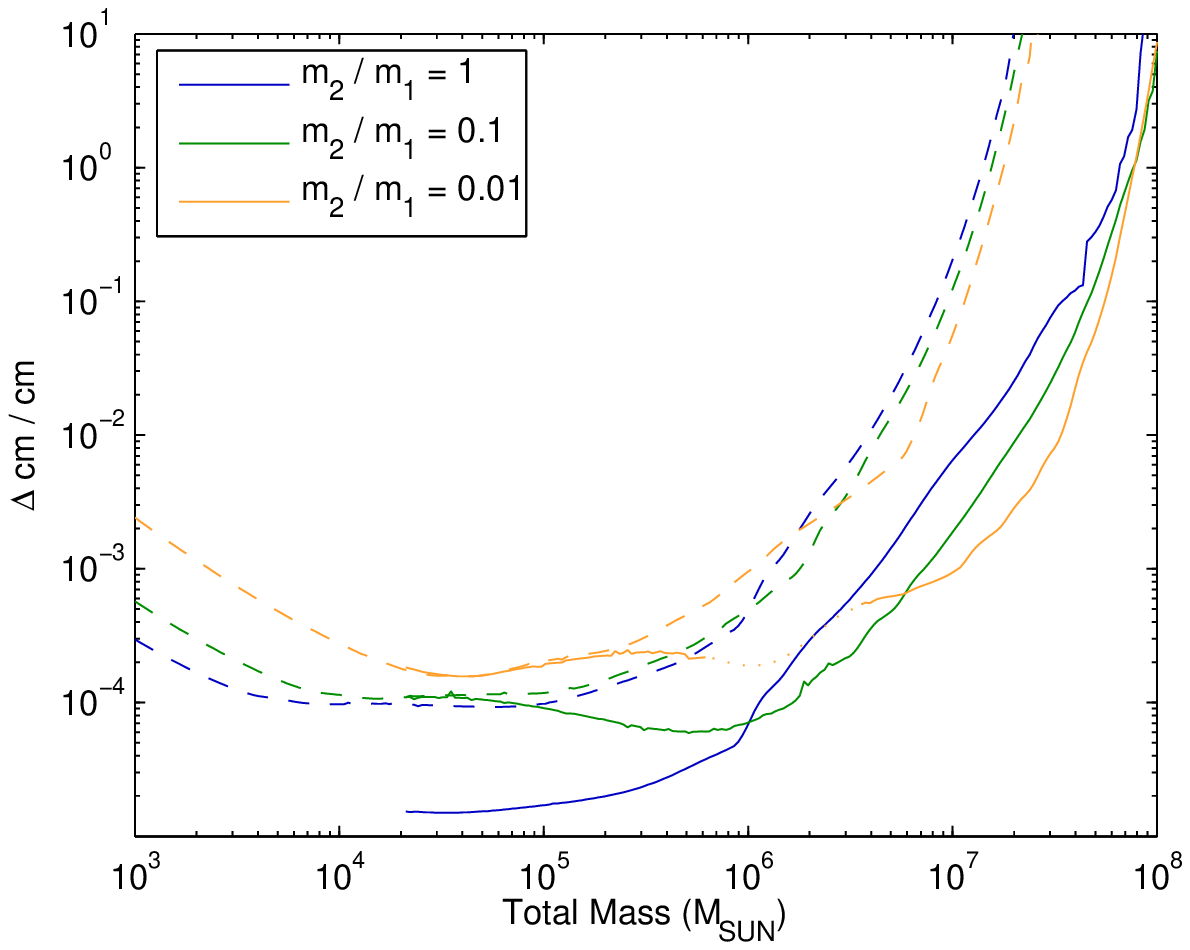} & \includegraphics[width=7.5cm]{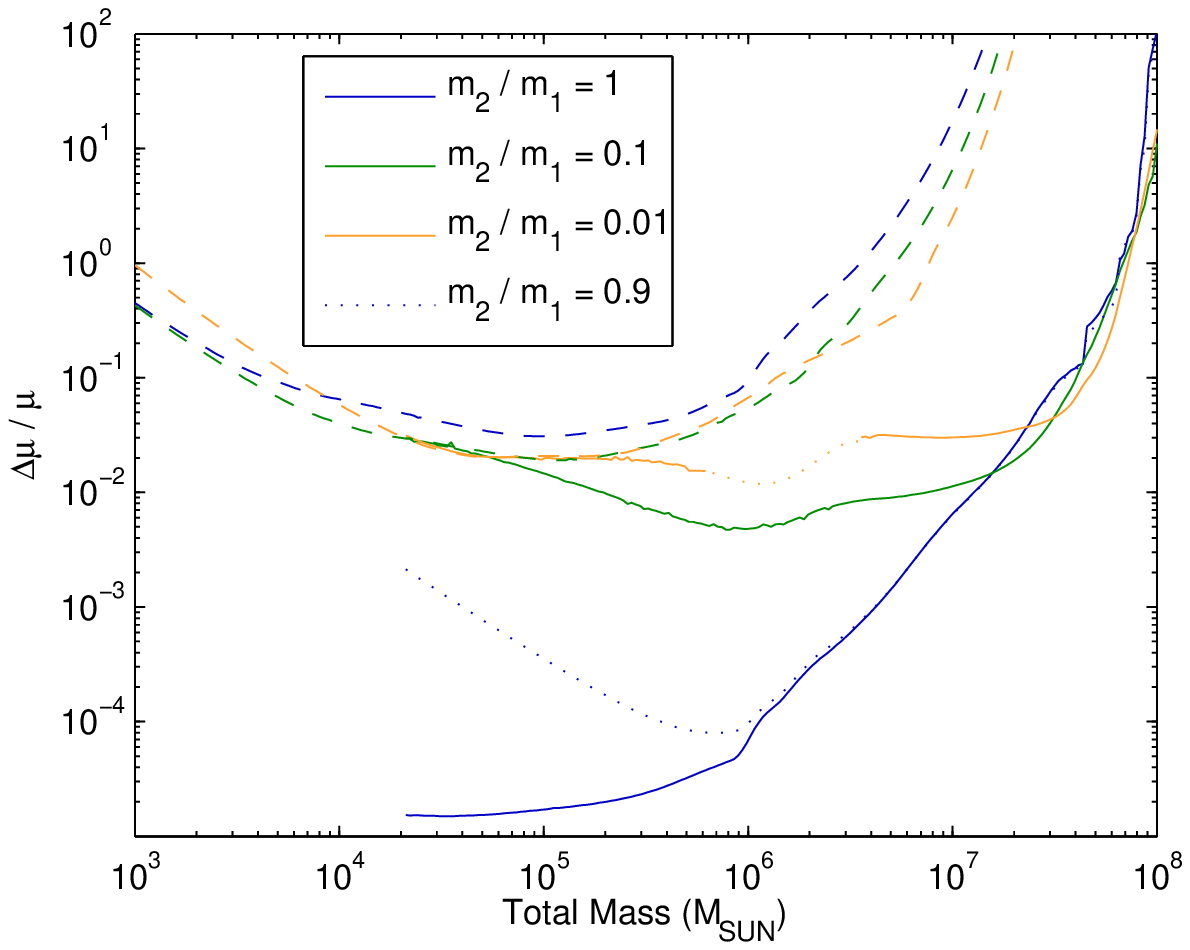}
\end{tabular}
\caption{Errors versus total mass for the same case as
 Fig.~\ref{Fig.FixedAng_SNR}.
 The solid lines correspond to FWF and dashed lines to RWF.}
\label{Fig.FixedAng_ERR}
\end{figure*}

\begin{figure}
 \includegraphics[width=8cm]{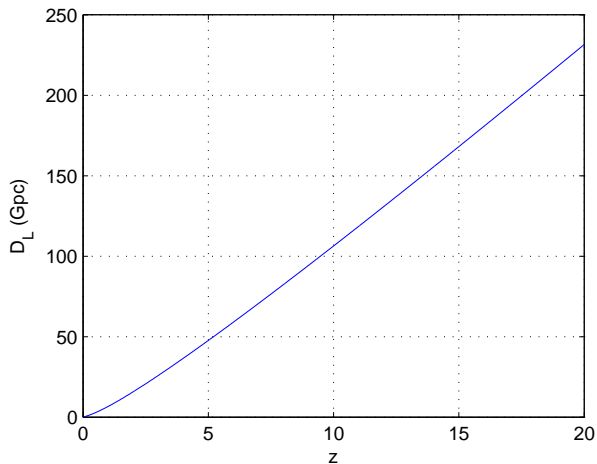} 
\caption{The luminosity distance as a function of redshift for a flat Universe described by 
 the cosmological parameters: 
 $H_0 = 71~\mathrm{km}~\mathrm{s}^{-1}~\mathrm{Mpc}^{-1}$,
 $\Omega_m = 0.27$ and $\Omega_\Lambda = 0.73$.}
\label{Fig.DLvsZ}
\end{figure}

\begin{figure}
\includegraphics[width=9cm]{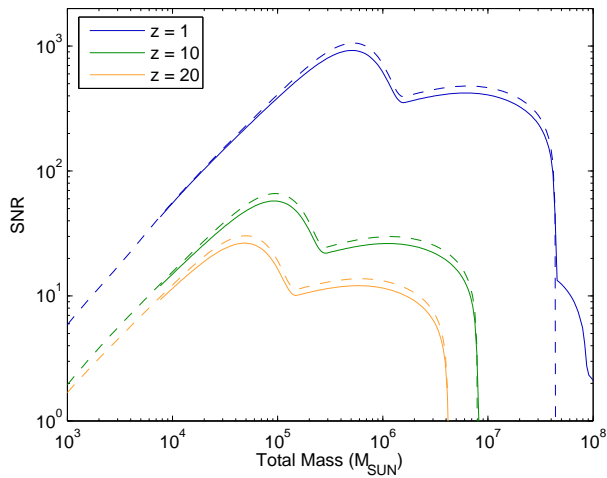}  
\caption{SNR versus total mass for various values of the source redshift.
We consider equal mass binary systems with fixed angles given by  
$\cos \theta_N = -0.6$, $\phi_N = 1$, $\cos \theta_L = 0.2$ and  
$\phi_L = 3$. The solid lines correspond the FWF and dashed lines to RWF.
}
\label{Fig.z_SNRs0}
\end{figure}

\begin{figure}
 \includegraphics[width=9cm]{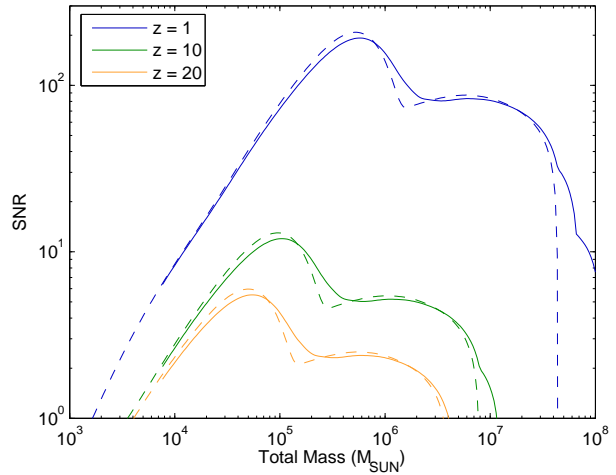} 
\caption{The same as Fig.~\ref{Fig.z_SNRs0} for a mass ratio $m_2/m_1 = 0.01$.
}
\label{Fig.z_SNRs}
\end{figure}

\begin{figure*}
\begin{tabular}{cc}
\includegraphics[width=8cm]{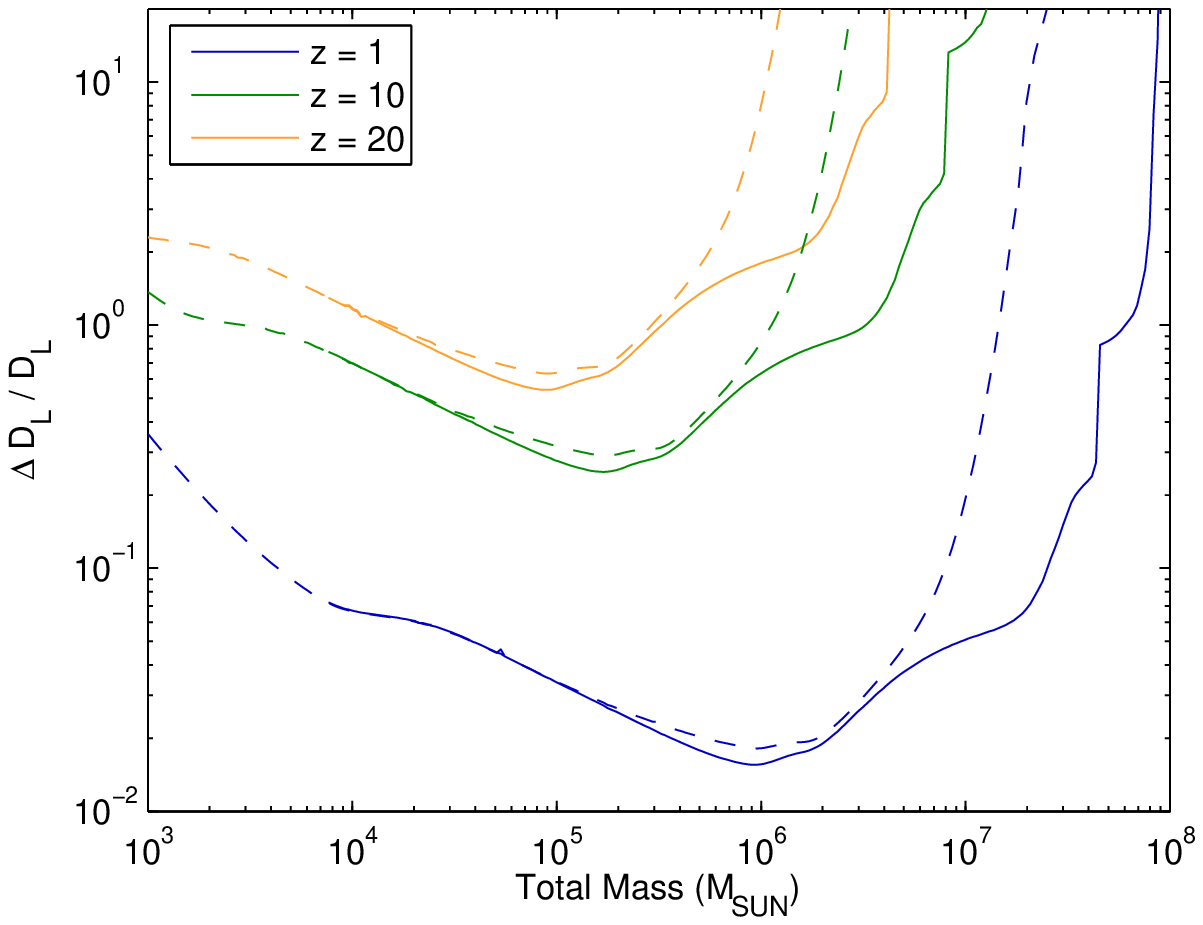} & \includegraphics[width=8cm]{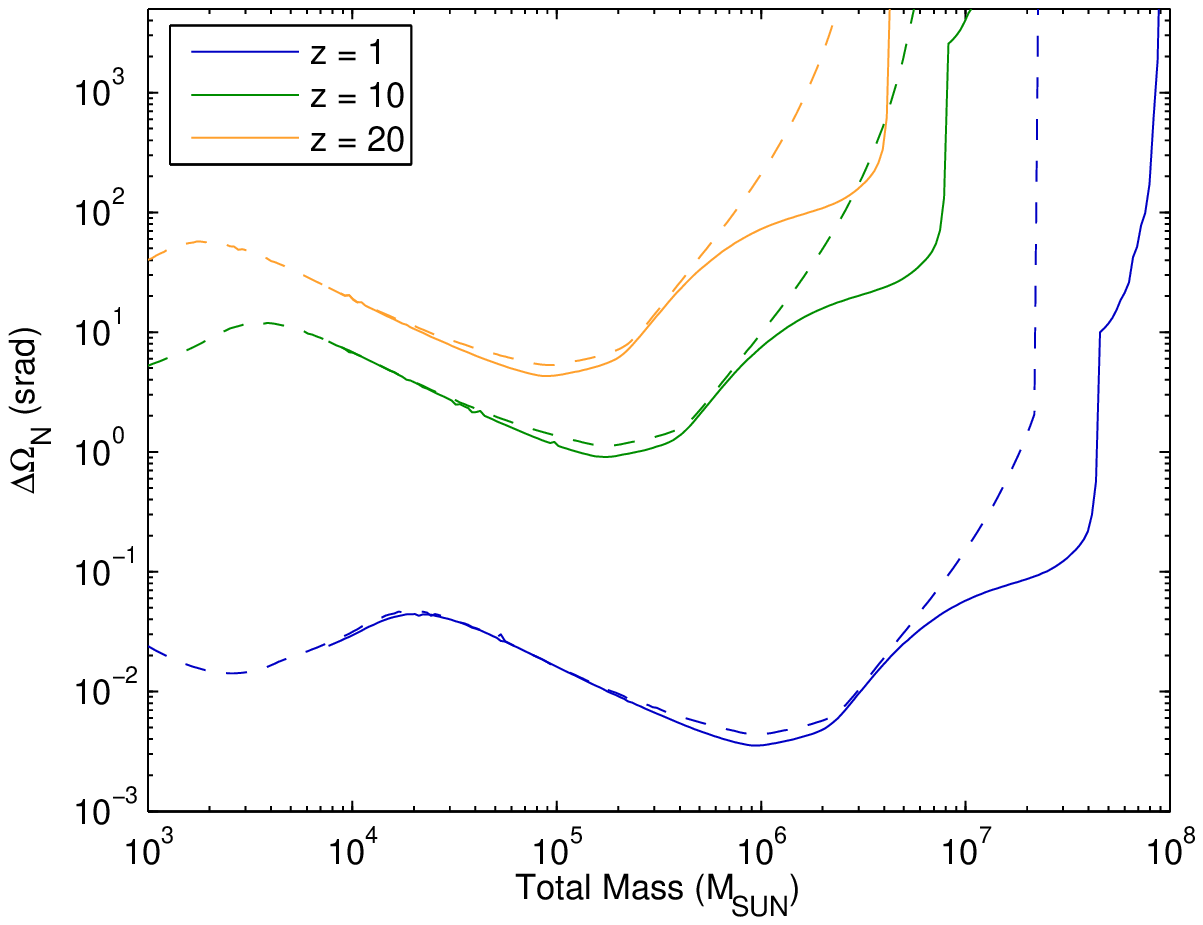} \\
\includegraphics[width=8cm]{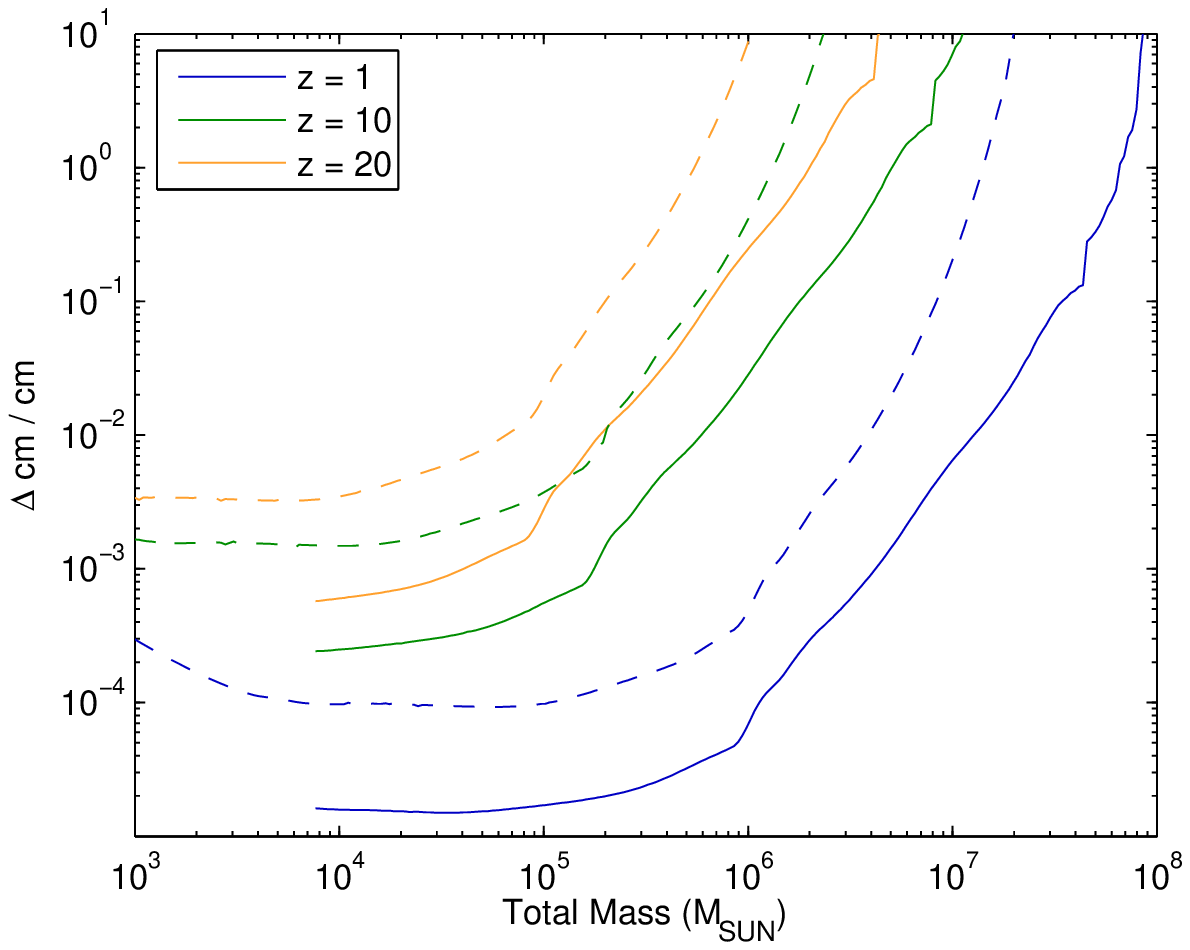} & \includegraphics[width=8cm]{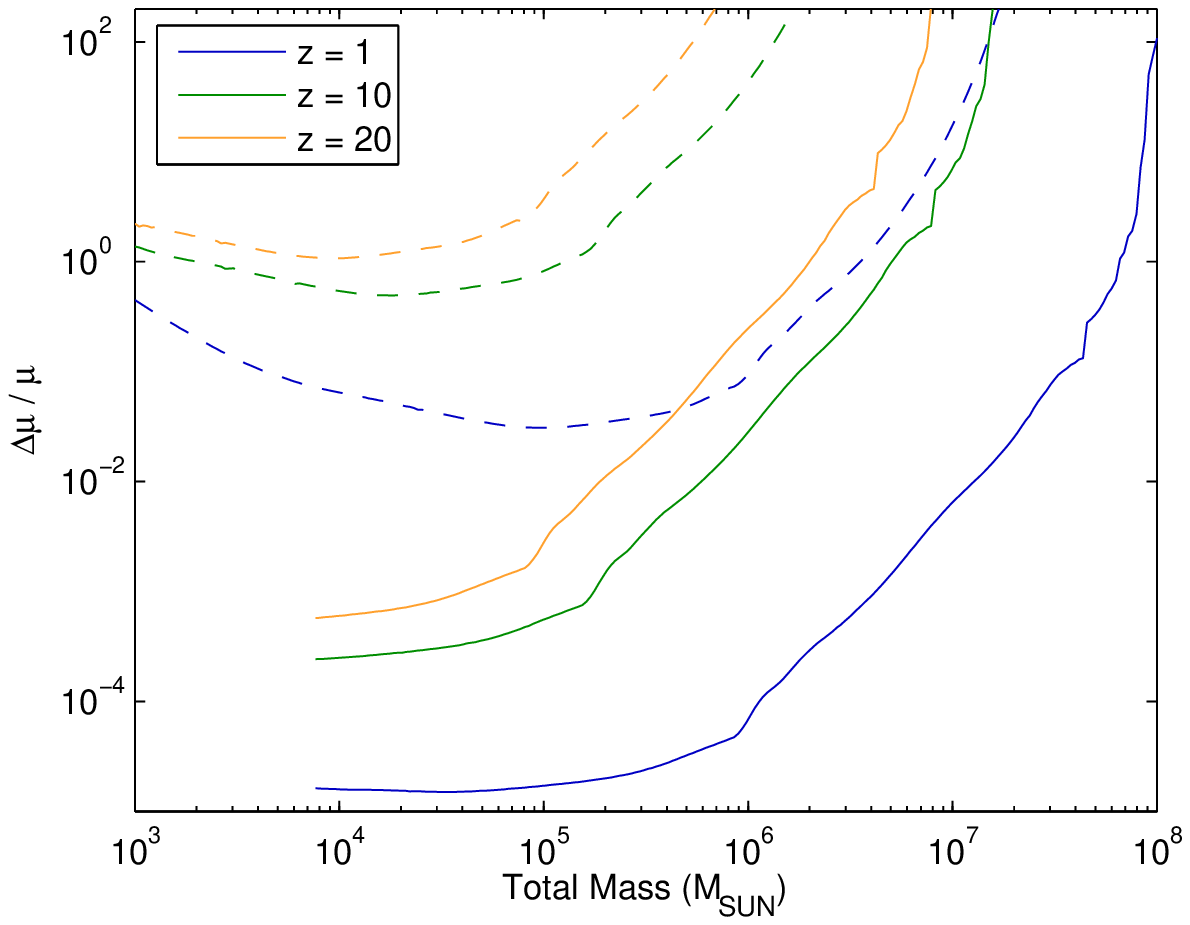}
\end{tabular}
\caption{
Distance measurement error, angular resolution, and mass measurement errors for 
LISA observations of the final year  of equal mass supermassive black hole inspirals.
The fiducial sources are at $z=$1, 10 and 20. The waveform considered are the FWF at 2PN order
(solid lines)
and the RWF (dashed lines) with $\cos \theta_N = -0.6$, $\phi_N = 1$, $\cos \theta_L = 0.2$,  
$\phi_L = 3$ and $\beta=\sigma=0$.
}
\label{Fig.z_ERR}
\end{figure*}

Given the extensive parameter space to be explored, we start by considering sources with a
fixed location and orientation given by $\cos \theta_N = -0.6$, $\phi_N = 1$, 
$\cos \theta_L = 0.2$ and  
$\phi_L = 3$. This is the same choice as in \cite{Arun:2007qv}, and, as we will point
out later, it corresponds to a case for which the SNR and the mass errors are similar to the 
typical average ones. Note that best and worst cases can span several orders of magnitude.
For this set of angles, we consider sources at redshift $z=1$, corresponding to a luminosity
distance of $D_L= 6.64$~Gpc and we study the effects on the SNR and parameter estimation
using the RWF and the FWF as function of the binary total mass.

In Fig.~\ref{Fig.FixedAng_SNR} we plot the SNR computed using the RWF and the FWF as 
a function
of the total mass of the binary system. For systems whose total mass $M<4\times 10^7 \Ms$ the
RWF over-estimates in general the SNR by a few percent. This was already pointed out 
in~\cite{Arun:2007qv}. It is also a known fact  that 
the Newtonian amplitude is about 7$\%$ higher than the 2.5PN order amplitude
and also than the amplitude obtained by numerical simulations over the last few 
orbits before merger.
For binaries with $M>4\times 10^7 \Ms$ the second harmonic 
is no longer visible in the LISA band
and higher harmonics, and therefore the use of FWF,
 play an important role extending the mass reach for supermassive black
holes. The 'jumps' at high masses are due to the low-frequency cut-off to the
instrument at $5\times 10^{-5}$~Hz.
In the case of the FWF we have limited our
study to systems with $M>2\times 10^4 \Ms$ due to our limited computational resources.

For the same configuration, in Fig.~\ref{Fig.FixedAng_ERR} we represent the errors
of the most relevant parameters as function of the total mass  for different mass ratios.
The Fisher matrix has been evaluated assuming the black hole spins to be zero, so that the 
spin-orbit and spin-spin parameters, $\beta$ and $\sigma$, respectively, are equal to
zero. In all cases, the errors are smaller for the FWF. For a total mass $M<10^5\Ms$  the improvements
are 
modest, except for the mass estimation for nearly equal masses, for which the errors in $\Mc$ and $\mu$
are of the same order; while these improvements are considerable for
 $M>5\times10^6\Ms$. This betterment is not due to an increase of SNR, but to  the higher
 harmonics that contribute to disentangle the source parameters. 
 For the equal mass case, the errors in $\mu$ improve up to 3 orders of magnitude 
 at $M=10^5\Ms$. In general the measurements of the masses improves by more than an order of magnitude
 for  $M>10^6\Ms$. 
 We find interesting to add the case $m_2/m_1=0.9$ obtained by the FWF 
 (dotted dark line, in the lower right panel) for comparison.
 Using the FWF the luminosity distance can be measured to better than
  $10\%$ up to $3-4\times 10^7\Ms$ for this particular source location,
  while using RWF the errors are
  several orders of magnitude larger for these large masses. Similarly, the error box in the sky improves
  significantly at $10^7\Ms$, although this sky position, as we will see below, was not one of the most
  favorable ones.
  
  Using the FWF and a mass ratio of $m_2/m_1=0.01$, there were several cases in which the errors we obtained
  did not have the desired accuracy. Those cases have been marked in Fig.~\ref{Fig.FixedAng_ERR} by
  replacing the solid line by a dotted one.
 
 LISA will also be able to measure gravitational waves from massive black hole coalescences to
 large redshift, making possible to study the merger history of black holes. Therefore, it is
 interesting to  extend the analysis to higher redshifts. 
 Figure \ref{Fig.DLvsZ} shows the luminosity distance as a function of redshift
 for a flat Universe described by 
 the cosmological parameters: 
 $H_0 = 71~\mathrm{km}~\mathrm{s}^{-1}~\mathrm{Mpc}^{-1}$,
 $\Omega_m = 0.27$ and $\Omega_\Lambda = 0.73$.
 In Figures \ref{Fig.z_SNRs0}
 and \ref{Fig.z_SNRs} we plot the SNR versus total mass for redshifts $z=1,10$ and 20.
 The results  we obtain are the expected ones since, modulo over-all amplitude, 
 the gravitational waves that we
 measure from a binary with masses $\{\Mc,\mu\}$ at redshift $z$  are those of a local
 system with masses  $\{(1+z)\Mc,(1+z)\mu\}$. The SNR decreases with $z$, not only due to the distance,
 but also because the frequency of the signal is redshifted and the total effective noise for LISA
 is higher at lower frequencies.

In Figure \ref{Fig.z_ERR} we plot the 
distance measurement error, angular resolution, and mass measurement errors for 
LISA observations of the final year of supermassive black hole inspirals.
The fiducial sources are at $z=$1, 10 and 20. The waveform considered are the FWF at 2PN order
and the RWF  with $\cos \theta_N = -0.6$, $\phi_N = 1$, $\cos \theta_L = 0.2$,  
$\phi_L = 3$ and $\beta=\sigma=0$.
For simplicity we display the curves corresponding to the equal mass case, but, as expected, other mass ratios  follow
the same trend. All curves drift with the redshift, but qualitatively the impact of FWF versus RWF is the same
as for $z=1$, but for different mass ranges.
 

\subsection{Exploring the parameter space}

\begin{figure*}
\begin{center}
\includegraphics[width=12cm]{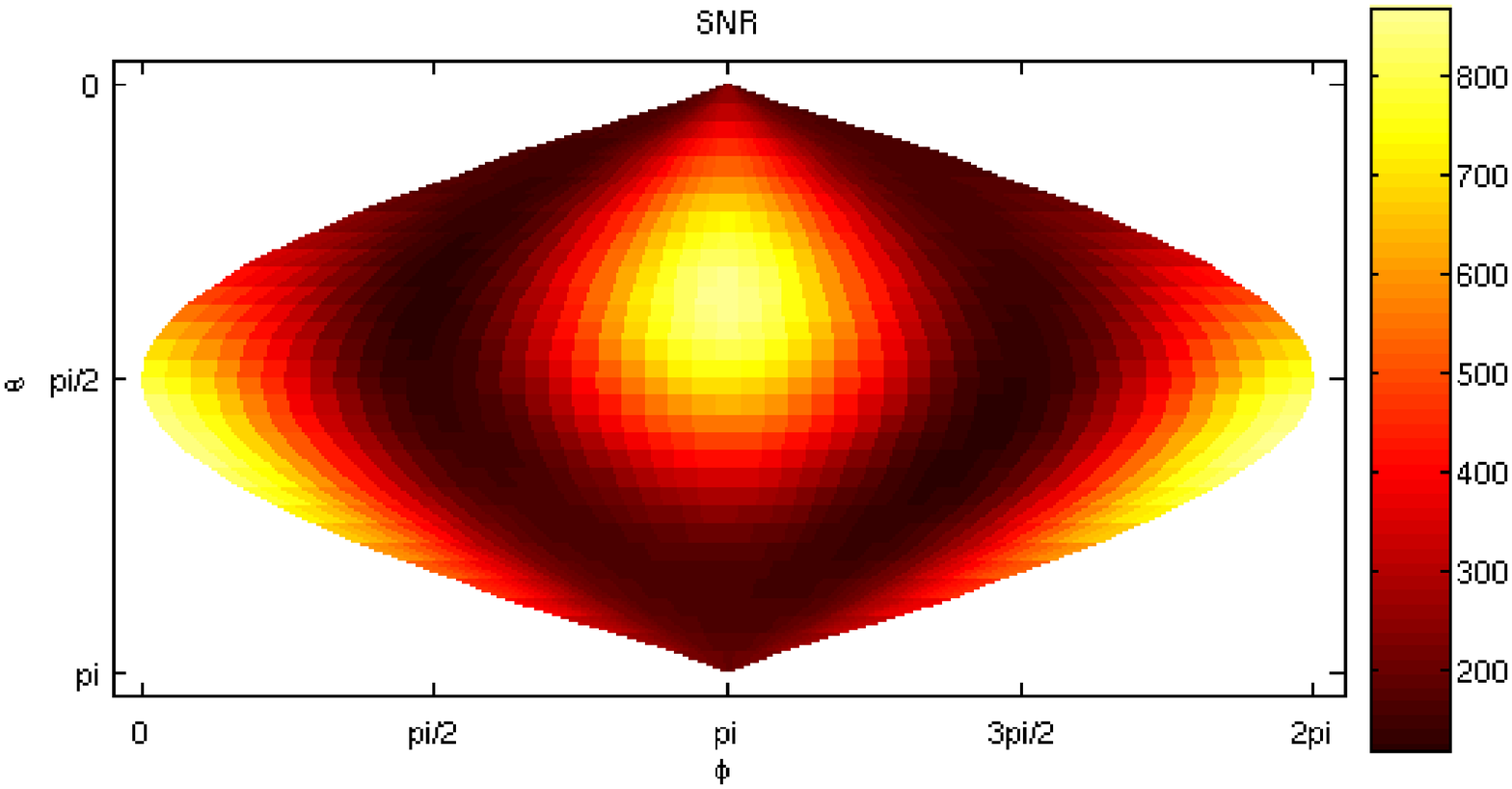}
\end{center}
\caption{ Sky map of SNR for LISA observations of the final year of inspirals using FWF.
The sources considered correspond to
$m_1 = m_2 = 10^7 \Ms$, at redshift $z=1$, with orientation angles
$\cos \theta_L = 0.2$ and $\phi_L = 3$. For all sources we assume the location of
 LISA at the time of coalescence is $\phi_{LISA} = 0$.}
\label{Fig.SkyMap_SNR}
\end{figure*}

\begin{figure*}
\begin{tabular}{cc}
\includegraphics[width=9cm]{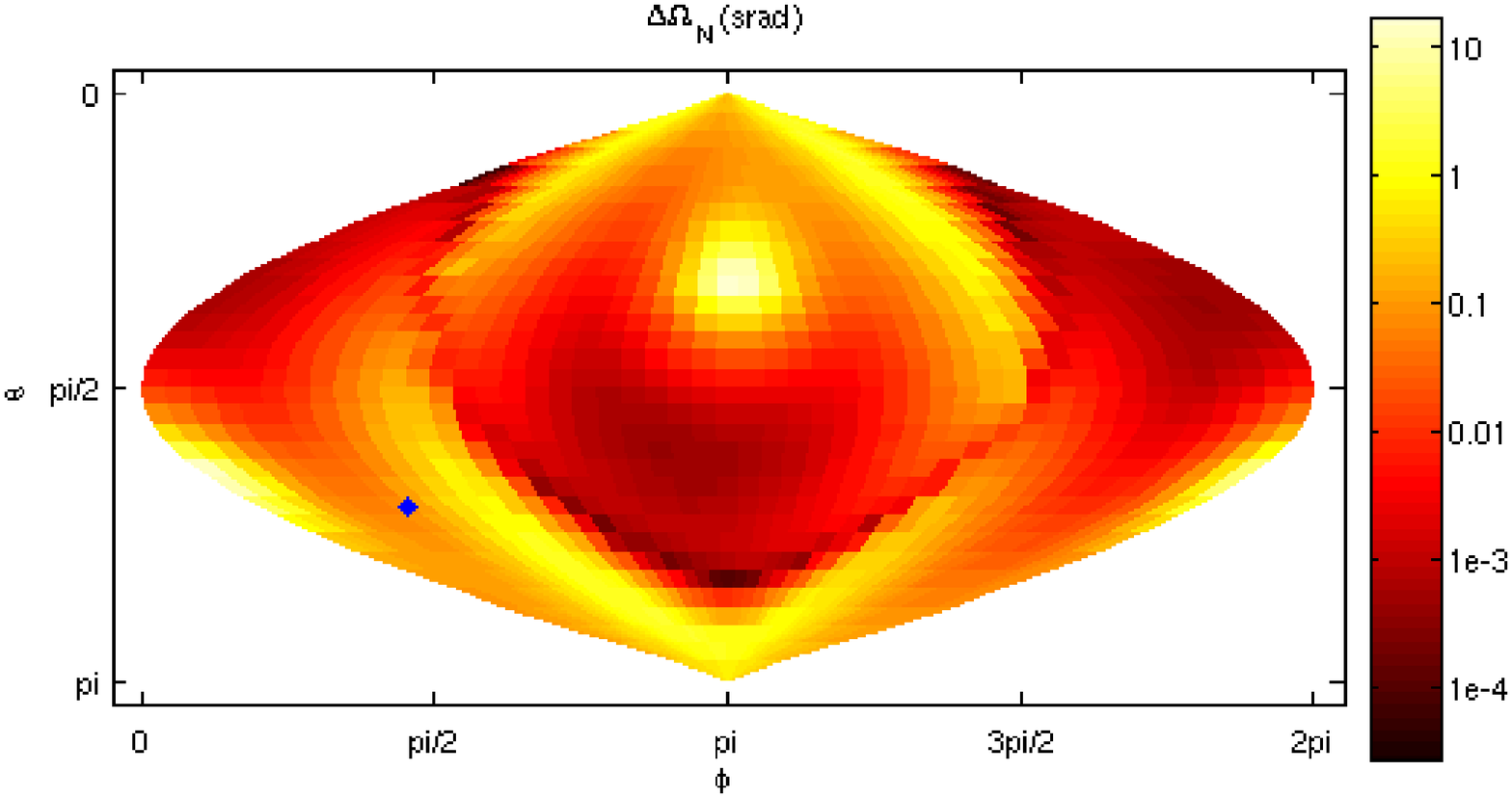} & \includegraphics[width=9cm]{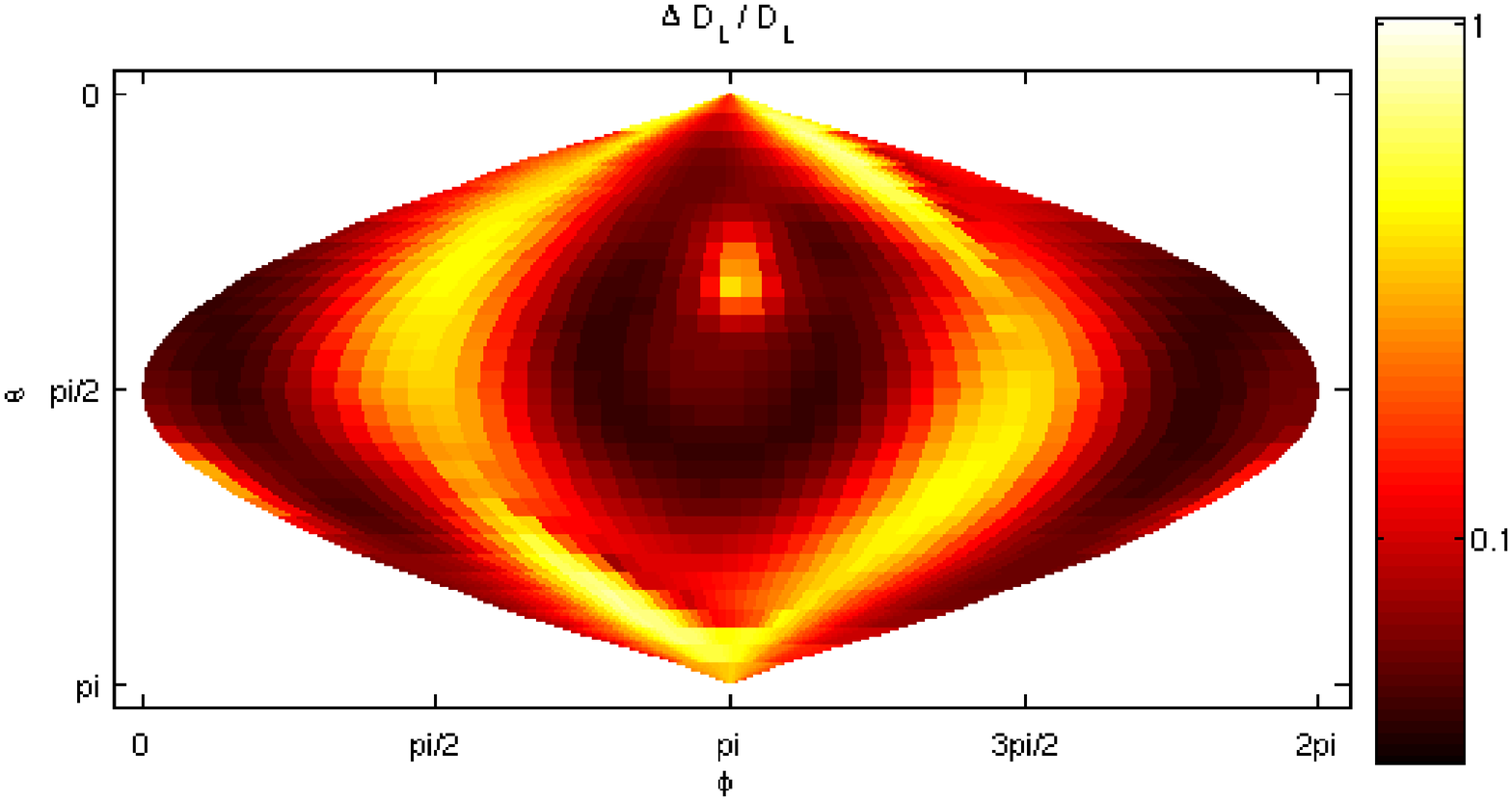} \\
\includegraphics[width=9cm]{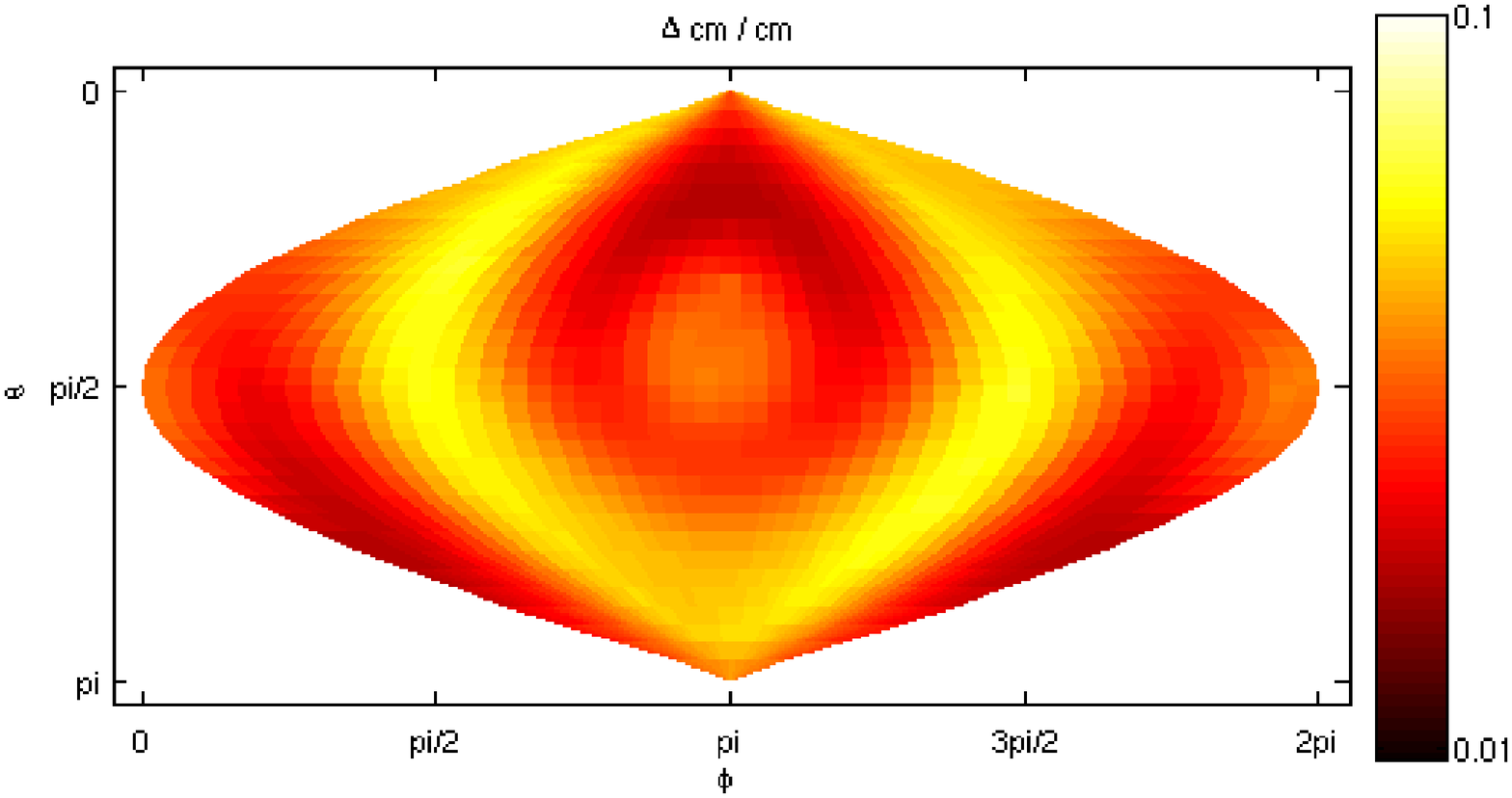} & \includegraphics[width=9cm]{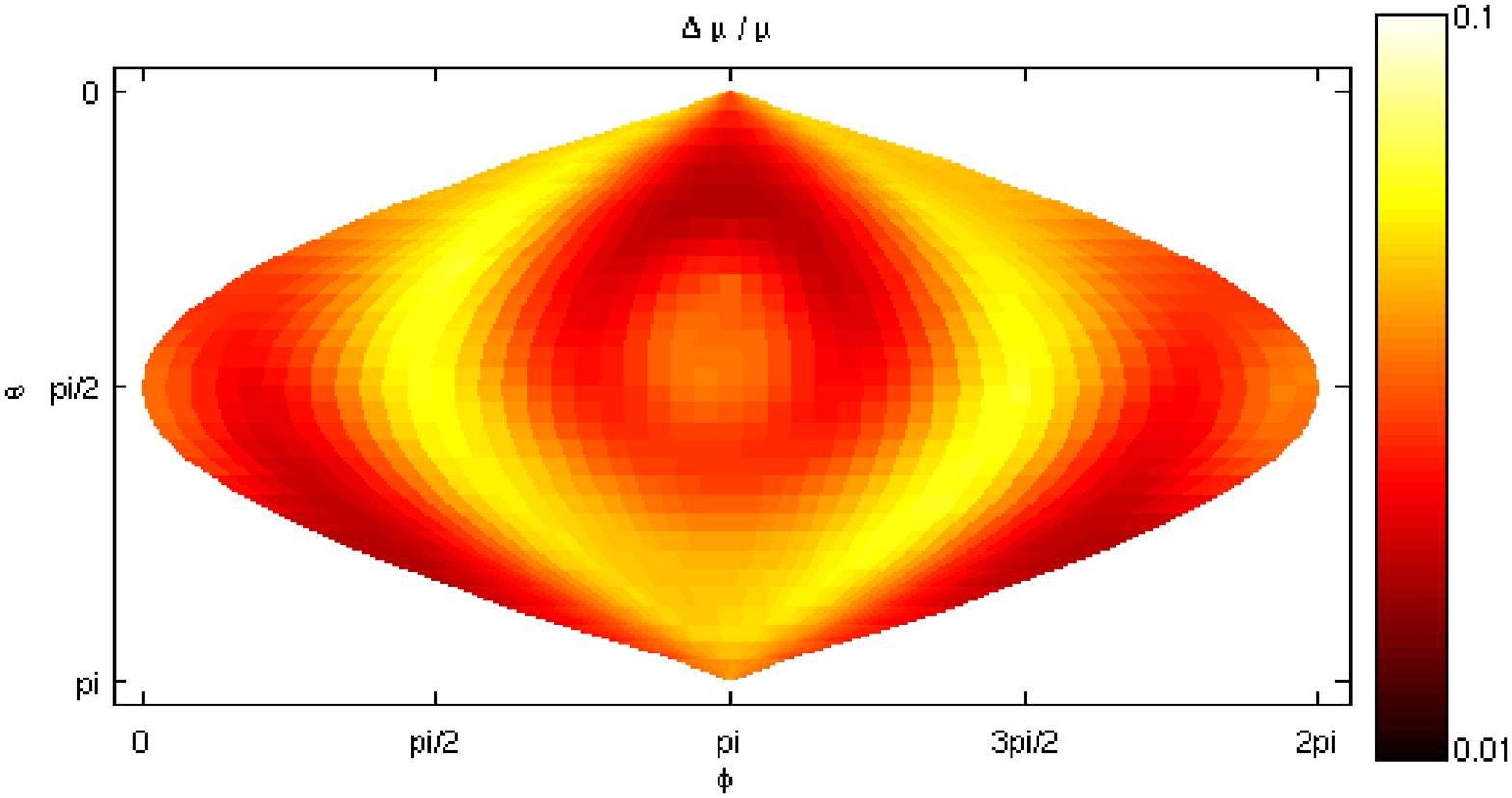}
\end{tabular}
\caption{Sky maps for the angular resolution, distance measurement error and mass measurement errors
for LISA observations of the final year of inspirals using FWF. As in Fig.~\ref{Fig.SkyMap_SNR}
these correspond to $m_1 = m_2 = 10^7 \Ms$ at redshift $z=1$, with orientation angles
$\cos \theta_L = 0.2$, $\phi_L = 3$ and $\beta=\sigma=0$. For all sources we assume 
the location of LISA at the time of coalescence is $\phi_{LISA} = 0$.
The dark blue dot corresponds to $\cos \theta_N = -0.6$, $\phi_N = 1$, mentioned in the text.}
\label{Fig.SkyMap_ERR}
\end{figure*}

\begin{figure*}

\includegraphics[width=12cm]{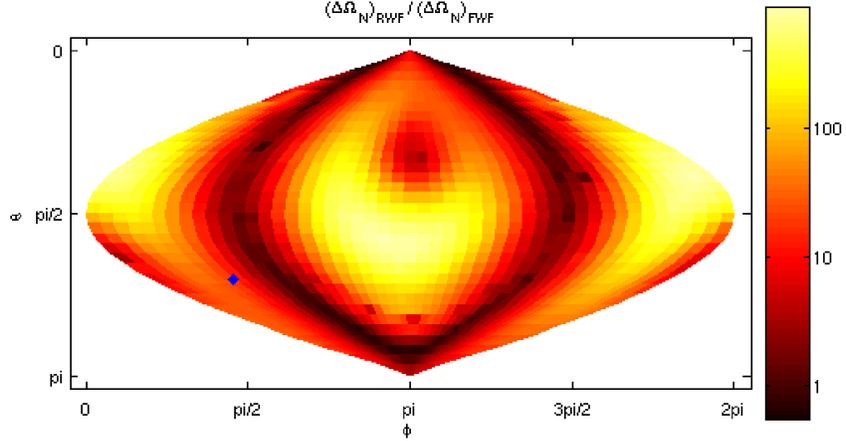} \\
\caption{Gain in angular resolution comparing the results obtained using FWF versus RWF
for the equal mass case  $m_1 = m_2 = 10^7 \Ms$ with the same assumptions as in 
Fig.~\ref{Fig.SkyMap_ERR}. The dark blue dot corresponds to $\cos \theta_N = -0.6$, $\phi_N = 1$.}
\label{Fig.SkyMap_GAIN77}
\end{figure*}

\begin{figure*}
\includegraphics[width=12cm]{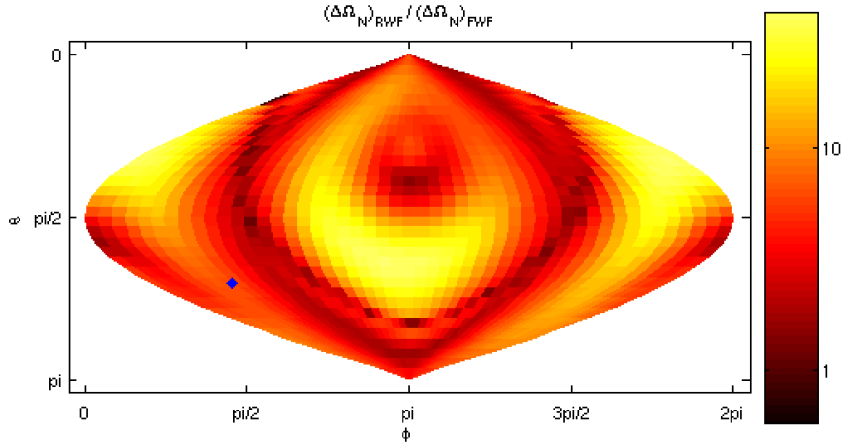} \\
\caption{Sky map of the gain in angular resolution for LISA 
observations of the final year of inspirals using FWF versus RWF corresponding to 
$m_1 = 10^7 M_{\odot}$, $m_2 = 10^6 M_{\odot}$ and $z=1$. 
We assume all source have the same orientation  
 $\cos \theta_L = 0.2$, $\phi_L = 3$, zero spins $\beta=\sigma=0$ and that LISA is  at
  $\phi_{LISA} = 0$ at the time of coalescence. 
  The dark blue dot corresponds to $\cos \theta_N = -0.6$, $\phi_N = 1$.
}
\label{Fig.SkyMap_GAIN}
\end{figure*}

\begin{figure*}
\begin{tabular}{cc}
(a) $m_1 = 10^7 M_{\odot}$ ; $m_2 = 10^7 M_{\odot}$  &  (b) $m_1 = 10^7 M_{\odot}$ ; $m_2 = 10^6 M_{\odot}$ \\
\includegraphics[width=9.0cm]{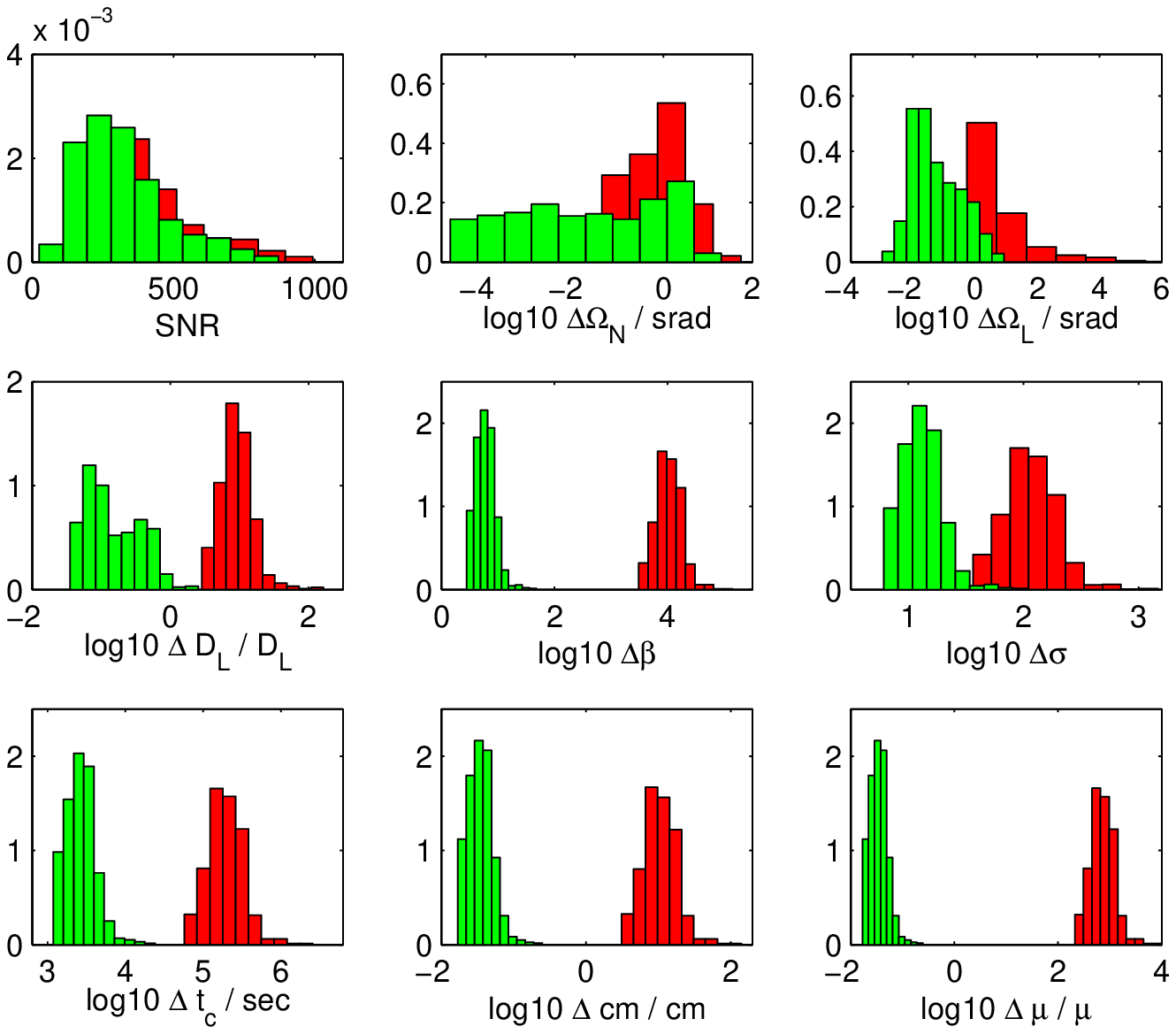} & \includegraphics[width=9.0cm]{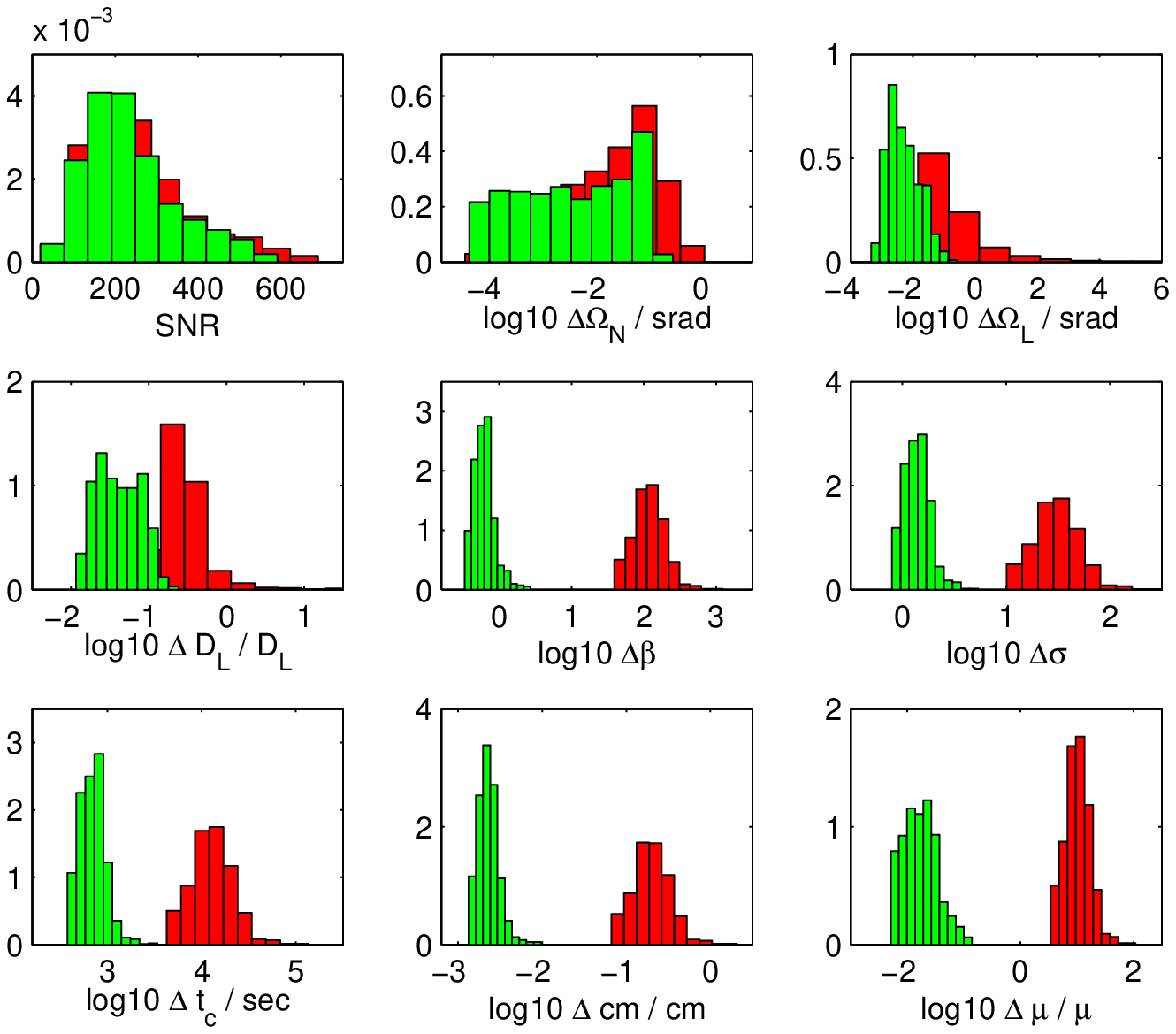} \\
(c) $m_1 = 10^7 M_{\odot}$ ; $m_2 = 10^5 M_{\odot}$  &  (d) $m_1 = 10^6 M_{\odot}$ ; $m_2 = 10^6 M_{\odot}$ \\
\includegraphics[width=9.0cm]{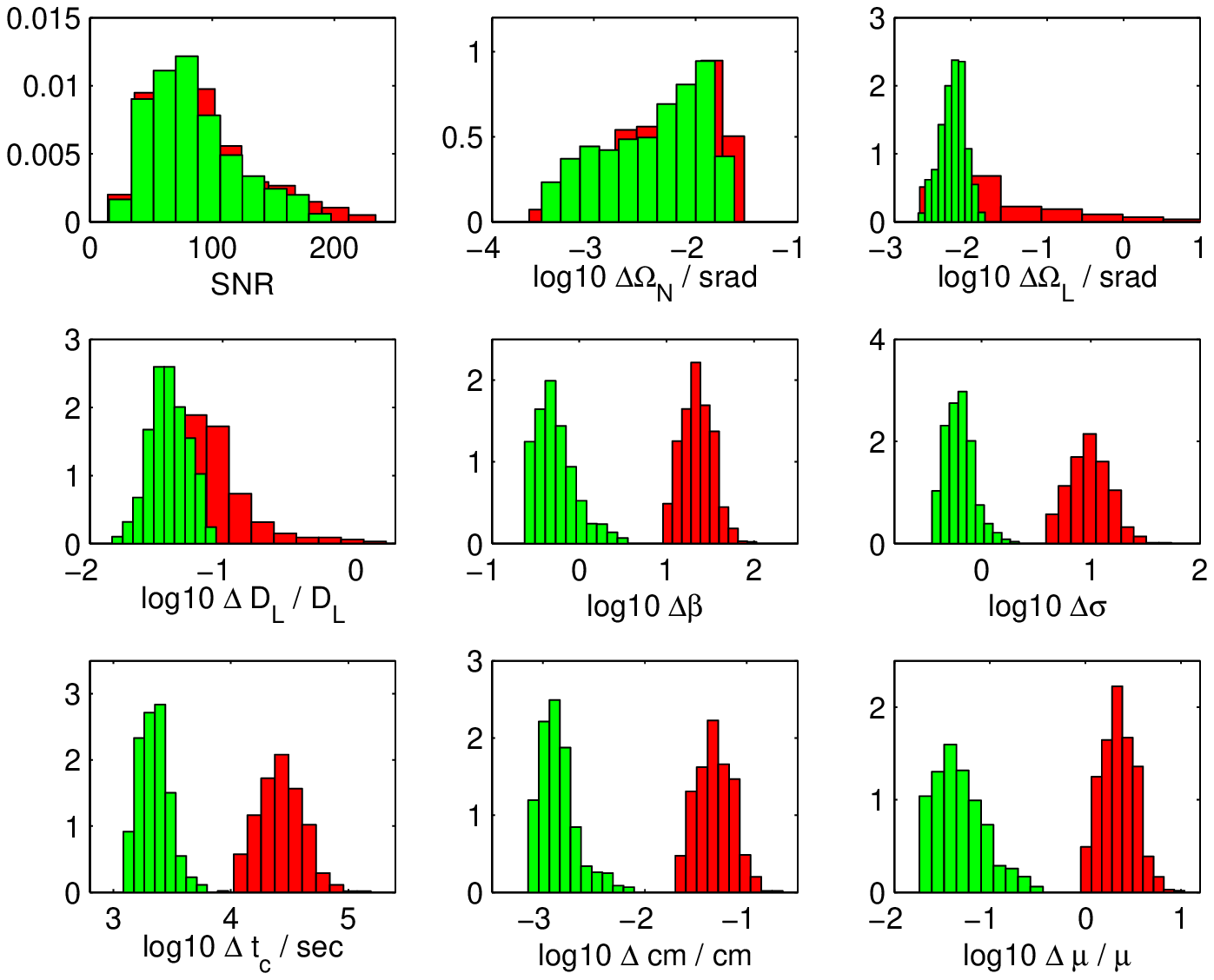} & \includegraphics[width=9.0cm]{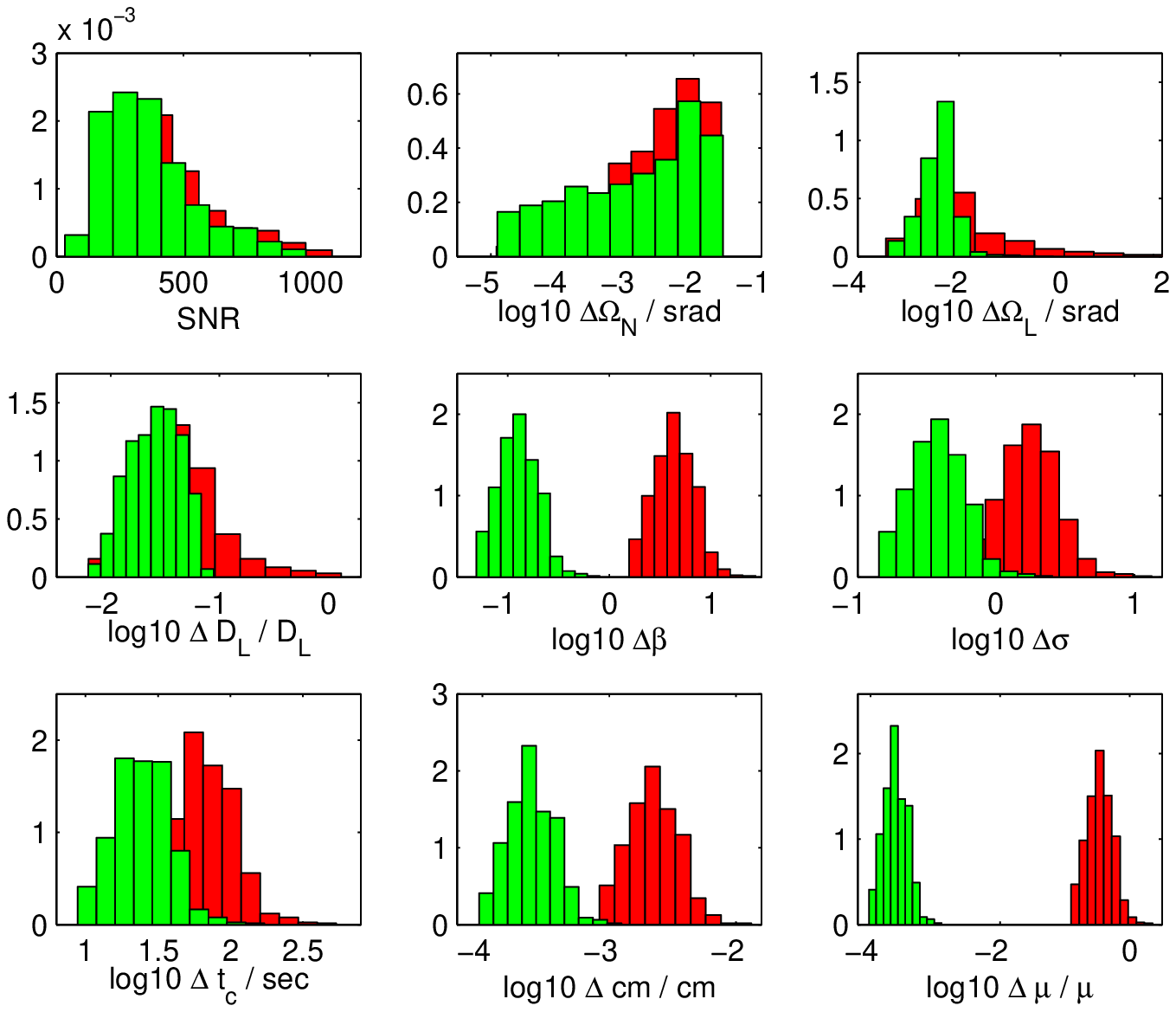} \\
(e) $m_1 = 10^6 M_{\odot}$ ; $m_2 = 10^5 M_{\odot}$  &  (f) $m_1 = 10^5 M_{\odot}$ ; $m_2 =
10^5 M_{\odot}$\\
\includegraphics[width=9.0cm]{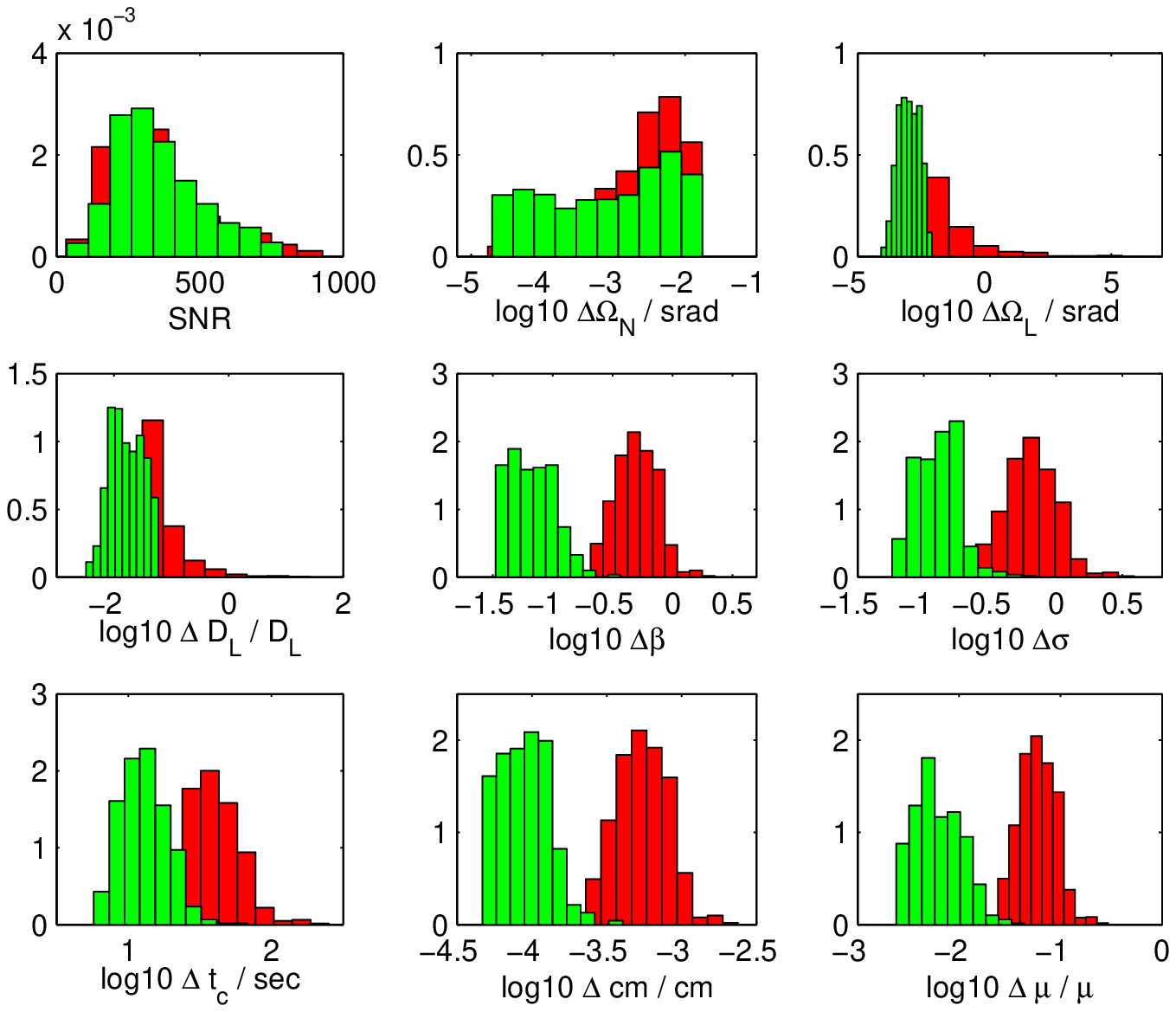} & \includegraphics[width=9.0cm]{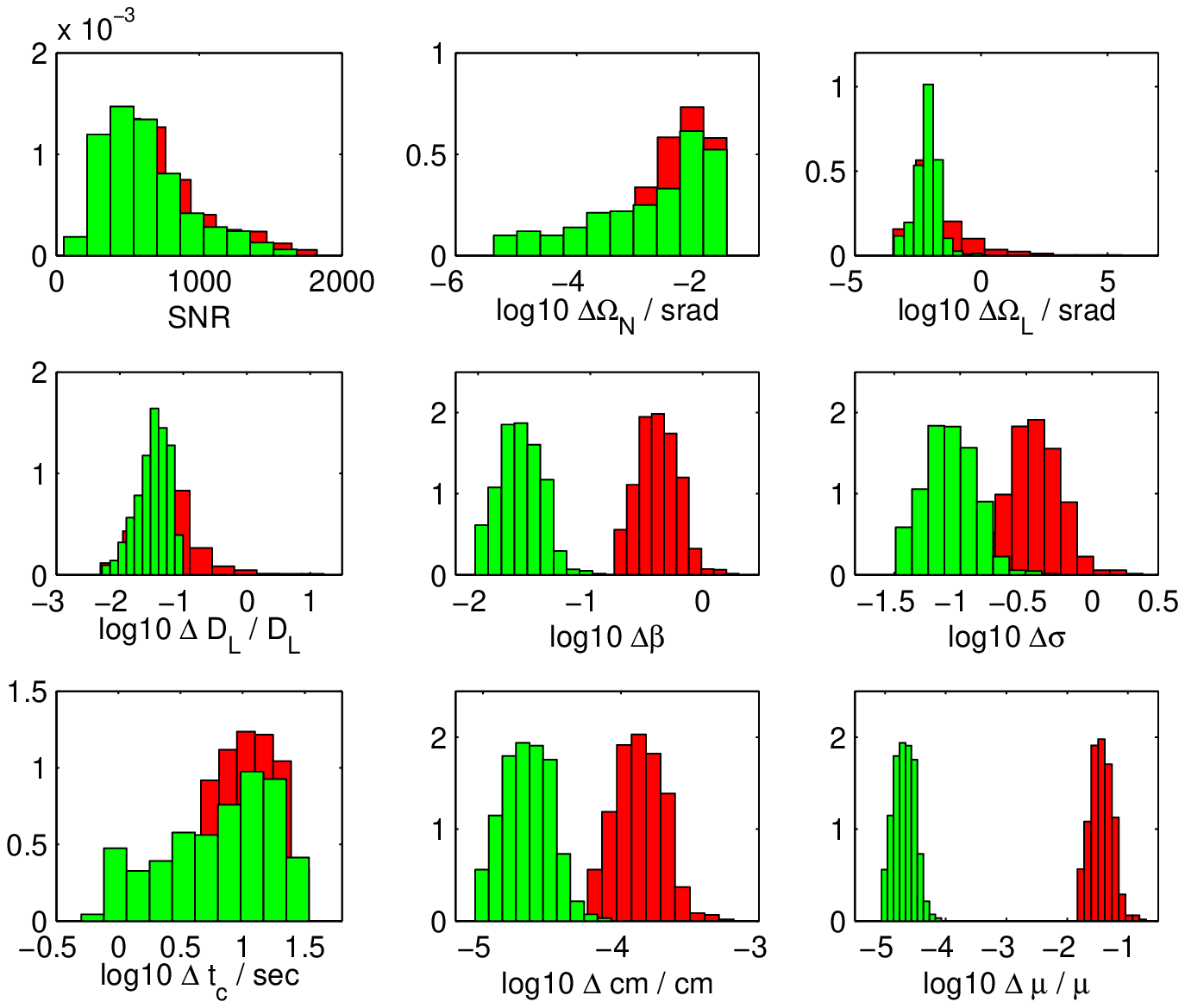} 
\end{tabular}
\caption{The probability distributions of SNR and measurement errors for observations of
the final year of supermassive black hole binaries at $z=1$ by LISA. For each pair of masses,
the histograms show the result of two Monte Carlo simulations, where 1000 sources have been
randomly located and oriented in the sky. The plots compare SNR and the errors
for the FWF (in clear green) and the RWF (in dark red).
}
\label{Fig.MCs}
\end{figure*}

\begin{table*}
\begin{center}
\begin{tabular}{|c|rrc|rrc|}
\hline 
& $\quad\quad\quad$  RWF$\quad$ & $\quad\quad\quad$  FWF$\quad$ & Gain factor & $\quad\quad\quad$  RWF$\quad$    & $\quad\quad\quad$  FWF$\quad$    &  Gain factor  \\
x&  $\langle x\rangle \pm \sigma_x$  &  $\langle x\rangle \pm \sigma_x$ &  & 
$\langle x\rangle \pm \sigma_x$  &  $\langle x\rangle \pm \sigma_x$   &    \\

\hline \hline 
& \multicolumn{3}{|c|}{\begin{normalsize} (a) $m_1 = 10^7 M_{\odot}$~;~$m_2 = 10^7 M_{\odot}$\end{normalsize}} &
  \multicolumn{3}{c|}{\begin{normalsize} (b) $m_1 = 10^7 M_{\odot}$~;~$m_2 = 10^6 M_{\odot}$\end{normalsize}} \\
\hline 
 SNR   & $370 \pm 183$ &  $322 \pm 160$ & $0.87$
                              & $258 \pm 127$ &  $237 \pm 110$ & $0.92$ \\
 $\log_{10} \Delta\Omega_N /$srad   & $-0.38 \pm 0.92$ &  $-1.78 \pm 1.67$ & $25$
                                                     & $-1.63 \pm 0.84$ &  $-2.50 \pm 1.07$ & $7.3$ \\
 $\log_{10} \Delta\Omega_L / $srad   & $0.40 \pm 1.20$ &  $-1.26 \pm 0.79$ & $46$
                                                     & $-0.84 \pm 1.20$ &  $-2.29 \pm 0.51$ & $28$ \\
 $\log_{10} \Delta D_L / D_L$   & $0.96 \pm 0.24$ &  $-0.83 \pm 0.39$ & $62$
                                                      & $-0.56 \pm 0.32$ &  $-1.39 \pm 0.27$ & $6.7$ \\
 $\log_{10} \Delta \beta$   & $4.02 \pm 0.23$ &  $0.78 \pm 0.18$ & $1750$
                                                  & $2.07 \pm 0.22$ &  $-0.21 \pm 0.14$ & $190$ \\
 $\log_{10} \Delta \sigma$   & $2.06 \pm 0.22$ &  $1.12\pm 0.18$ & $8.8$
                                                   & $1.48 \pm 0.22$ &  $0.14 \pm 0.12$ & $22$ \\
 $\log_{10} \Delta t_c / $s   & $5.28 \pm 0.23$ &  $3.42 \pm 0.19$ & $72$
                                                      & $4.11 \pm 0.22$ &  $2.84 \pm 0.14$ & $18$ \\
 $\log_{10} \Delta \cm / \cm$   & $1.02 \pm 0.23$ &  $-1.46 \pm 0.17$ & $300$
                                                      & $-0.70 \pm 0.22$ &  $-2.64 \pm 0.12$ & $87$ \\
 $\log_{10} \Delta \mu / \mu$   & $2.85 \pm 0.23$ &  $-1.46 \pm 0.17$ & $20000$
                                                      & $1.01 \pm 0.22$ &  $-1.77 \pm 0.30$ & $600$ \\
\hline
& \multicolumn{3}{|c|}{\begin{normalsize} (c) $m_1 = 10^7 M_{\odot}$~;~$m_2 = 10^5 M_{\odot}$\end{normalsize}} &
  \multicolumn{3}{c|}{\begin{normalsize}(d) $m_1 = 10^6 M_{\odot}$~;~$m_2 = 10^6 M_{\odot}$\end{normalsize}} \\
\hline  
 SNR   & $90 \pm 42$ &  $85 \pm 36$ & $0.94$ 
                              & $405 \pm 200$ &  $365 \pm 181$ & $0.90$ \\
 $\log_{10} \Delta\Omega_N / $srad   & $-2.29 \pm 0.48$ &  $-2.42 \pm 0.49$ & $1.4$
                                                     & $-2.58 \pm 0.70$ &  $-2.89 \pm 0.93$ & $2.1$ \\
 $\log_{10} \Delta\Omega_L / $srad   & $-1.50 \pm 1.01$ &  $-2.23 \pm 0.16$ & $5.4$
                                                     & $-1.85 \pm 1.02$ &  $-2.40 \pm 0.36$ & $3.5$ \\
 $\log_{10} \Delta D_L / D_L$   & $-1.05 \pm 0.27$ &  $-1.40 \pm 0.14$ & $2.3$
                                                      & $-1.34 \pm 0.37$ &  $-1.60 \pm 0.23$ & $1.8$ \\
 $\log_{10} \Delta \beta$   & $1.35 \pm 0.18$ &  $-0.27 \pm 0.23$ & $41$
                                                  & $0.64 \pm 0.20$ &  $-0.90 \pm 0.20$ & $34$ \\
 $\log_{10} \Delta \sigma$   & $0.99 \pm 0.18$ &  $-0.20 \pm 0.13$ & $15$
                                                   & $0.24 \pm 0.20$ &  $-0.42 \pm 0.20$ & $4.6$ \\
 $\log_{10} \Delta t_c / $s   & $4.42 \pm 0.18$ &  $3.35 \pm 0.13$ & $12$
                                                      & $1.84 \pm 0.19$ &  $1.39 \pm 0.19$ & $2.8$ \\
 $\log_{10} \Delta \cm / \cm$   & $-1.31 \pm 0.18$ &  $-2.85 \pm 0.18$ & $34$
                                                      & $-2.65 \pm 0.19$ &  $-3.61 \pm 0.18$ & $9.1$ \\
 $\log_{10} \Delta \mu / \mu$   & $0.34 \pm 0.18$ &  $-1.31 \pm 0.27$ & $46$
                                                      & $-0.46 \pm 0.20$ &  $-3.61 \pm 0.18$ & $1420$ \\
\hline
& \multicolumn{3}{|c|}{\begin{normalsize}(e) $m_1 = 10^6 M_{\odot}$~;~$m_2 = 10^5 M_{\odot}$\end{normalsize}} &
  \multicolumn{3}{c|}{\begin{normalsize}(f) $m_1 = 10^5 M_{\odot}$~;~$m_2 = 10^5 M_{\odot}$\end{normalsize}} \\
\hline 
 SNR   & $348 \pm 170$ &  $356 \pm 152$ & $1.0$
                              & $680 \pm 640$ &  $620 \pm 310$ & $0.91$ \\
 $\log_{10} \Delta\Omega_N / $srad   & $-2.64 \pm 0.63$ &  $-3.10 \pm 0.87$ & $2.9$
                                                     & $-2.45 \pm 0.69$ &  $-2.80 \pm 1.00$ & $2.2$ \\
 $\log_{10} \Delta\Omega_L / $srad   & $-1.82 \pm 1.22$ &  $-2.98 \pm 0.41$ & $15$
                                                     & $-1.66 \pm 1.21$ &  $-2.14 \pm 0.48$ & $3.0$ \\
 $\log_{10} \Delta D_L / D_L$   & $-1.35 \pm 0.44$ &  $-1.78 \pm 0.29$ & $2.7$
                                                      & $-1.26 \pm 0.45$ &  $-1.48 \pm 0.25$ & $1.7$ \\
 $\log_{10} \Delta \beta$   & $-0.30 \pm 0.17$ &  $-1.17 \pm 0.19$ & $7.4$
                                                  & $-0.40 \pm 0.18$ &  $-1.62 \pm 0.19$ & $17$ \\
 $\log_{10} \Delta \sigma$   & $-0.19 \pm 0.19$ &  $-0.89 \pm 0.16$ & $5.1$
                                                   & $-0.44 \pm 0.19$ &  $-1.07 \pm 0.20$ & $4.4$ \\
 $\log_{10} \Delta t_c / $s   & $1.57 \pm 0.19$ &  $1.11 \pm 0.16$ & $2.9$
                                                      & $0.97 \pm 0.29$ &  $0.80 \pm 0.44$ & $1.5$ \\
 $\log_{10} \Delta \cm / \cm$   & $-3.26 \pm 0.17$ &  $-4.05 \pm 0.16$ & $6.1$
                                                      & $-3.86 \pm 0.17$ &  $-4.68 \pm 0.17$ & $6.6$ \\
 $\log_{10} \Delta \mu / \mu$   & $-1.22 \pm 0.18$ &  $-2.21 \pm 0.23$ & $9.6$
                                                      & $-1.44 \pm 0.18$ &  $-4.68 \pm 0.17$ & $1760$ \\
\hline
\end{tabular}
\caption{Characterization of the probability distributions of SNR and measurement errors
of Fig.~\ref{Fig.MCs}. For each pair of masses and waveform model used, the mean and
standard deviation of the SNR is given, as well as the mean and standard deviation of the logarithm
of the measurement errors. The gain factors are computed as  
$\langle SNR\rangle_{\FWF}/ \langle SNR\rangle_{\RWF}$
and  $ 10^{(\langle x\rangle_{\RWF}-\langle x\rangle_{\FWF})}$, for the SNR and measurement errors,
respectively.}
\label{Tab.MCs}
\end{center}
\end{table*}

\begin{figure}
 \includegraphics[width=9cm]{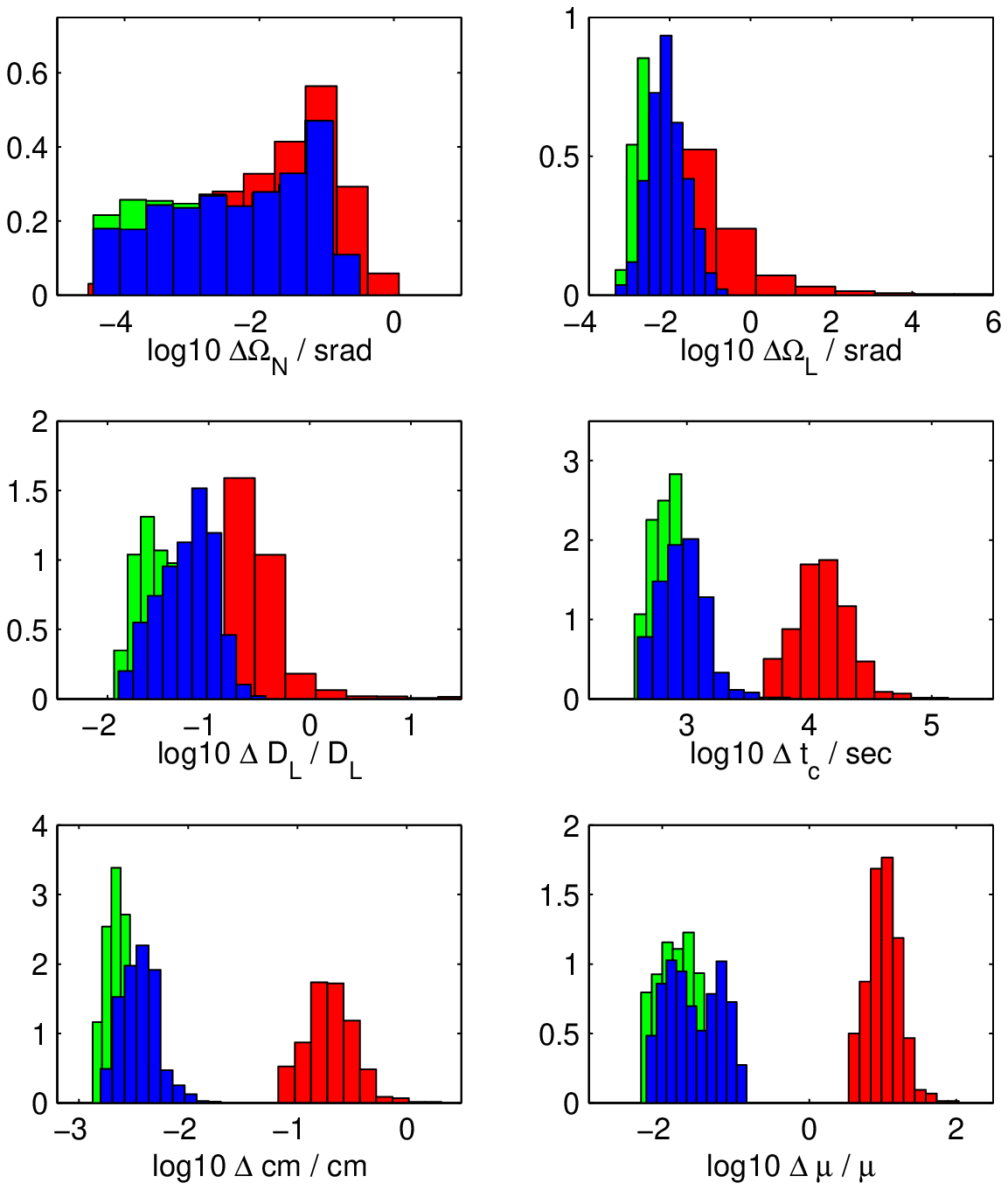} 
\caption{Comparison of measurement errors for the FWF (in clear green), the RWF (in dark red)
and a waveform containing only the second and third harmonics (in very dark blue),
 corresponding to LISA 
observations of the final year of inspirals for
$m_1 = 10^7 M_{\odot}$, $m_2 = 10^6 M_{\odot}$ and $z=1$. 
}
\label{Fig.hist76semi}
\end{figure}

As we mentioned in the previous section, the error measurements are very sensitive 
-- they vary by orders-of-magnitude -- to the true value of the source parameters; 
in order to
give meaningful results, one therefore is forced to explore a large parameter space.
We do this
(i) by considering sources on an isotropic grid in the sky with a fixed orientation,
and (ii) by extensive Monte-Carlo simulations for all possible location and orientation of the source
with respect to LISA.

Figure \ref{Fig.SkyMap_SNR} is a sample of a
sky map of SNR for LISA observations of the final year of inspirals using FWF.
The sources considered correspond to
$m_1 = m_2 = 10^7 \Ms$ at redshift $z=1$, with orientation angles
$\cos \theta_L = 0.2$ and $\phi_L = 3$. 
For all sources we assume the location of
 LISA at the time of coalescence is $\phi_{LISA} = 0$.
 The SNR over the entire sky covers a
range $\sim 100-900$. The SNR is higher for sources located orthogonal to the plane of
LISA at the time of coalescence, which corresponds to $(\phi_N=0\oo; \theta_N=120\oo)$
and $(\phi_N=180\oo; \theta_N=60\oo)$,
or in general
$(\phi_{\bot} = \phi_{LISA} ; \theta_{\bot} = \theta_{LISA} + 30\oo)$ and
$(\phi_{\bot} = \phi_{LISA} + 180\oo ; \theta_{\bot} = \theta_{LISA} - 30\oo)$,
since the LISA constellation is inclined at an angle of $60\oo$ with respect 
to the ecliptic.
 The reason is that most of the SNR is accumulated in the last days before merger.
The reader can notice also the nearly symmetric 
$[(\theta, \phi)\rightarrow (\pi-\theta,\phi+\pi)]$ profiles of 
figures \ref{Fig.SkyMap_SNR}-\ref{Fig.SkyMap_GAIN}. 
The small asymmetry shows the relative importance of the Doppler phase modulation.

Figure \ref{Fig.SkyMap_ERR} shows the angular resolution, 
and the error in the measurements of the luminosity distance and masses
over the entire sky for the same case as Fig.~\ref{Fig.SkyMap_SNR}. To summarize, the parameters
of two $10^7\Ms$ black holes spiraling toward the final merger at $z=1$ can be measured very
accurately depending on the sky location: $\Delta\Omega_N$ up to $\sim 10^{-4}$~srad, 
the luminosity distance to better than $1\%$, and the masses between $1\%$ and $10\%$.
Although it is true that for a given source, the higher SNR the better the parameter estimation, 
e.g., by changing the distance, this cannot be generalized comparing the SNR at different locations
in the sky. It is not just a matter of SNR but long observation times that contribute to
disentangle and improve the parameter estimation.

In Fig.~\ref{Fig.SkyMap_GAIN77} we show the gain in angular resolution comparing the results obtained 
using FWF versus RWF for the  case  $m_1 = m_2 = 10^7 \Ms$  as in 
Fig.~\ref{Fig.SkyMap_ERR}. The benefit of using FWF is clear. For all those sky locations in which
we obtain the best angular resolution,  $\Delta\Omega_N\sim 10^{-4}$~srad with the FWF, the 
corresponding  gain is up to 3 orders of magnitude.

Figure \ref{Fig.SkyMap_GAIN} represents also the improvement
 in angular resolution for the unequal mass
case $m_1 = 10^7 \Ms$, $m_2 = 10^6 \Ms$. In this case the optimal gain is of $\sim 2$ orders of
magnitude, and those correspond to the same sky locations as in the equal mass case analyzed before.

In order to cover completely the full parameter space, 
we proceed to perform
 Monte-Carlo simulations according to the following: 
we consider an
ensemble of fiducial sources all at redshift $z=1$ (which sets the luminosity distance 
$D_L=6.64$~Gpc),
with zero spins $\beta=\sigma=0$
and   we select the value of the masses $m_1$ and $m_2$. 
For each set of mass parameters
we select randomly the four geometrical angles ($\theta_N$, $\phi_N$, $\theta_L$ and
$\phi_L$) from an uniform distribution in 
$\cos\theta_N$, $\phi_N$, $\cos\theta_L$ and $\phi_L$, and 
 as far as for the other two parameters, we chose them as
$t_c=\phi_c=0$.
The Monte-Carlo is done on a 1000 different sets of angles.
We present the results in terms of probability distributions.

We have studied six different pairs of masses:
(a) $m_1 = 10^7 M_{\odot}$; $m_2 = 10^7 M_{\odot}$,
(b) $m_1 = 10^7 M_{\odot}$; $m_2 = 10^6 M_{\odot}$, 
(c) $m_1 = 10^7 M_{\odot}$; $m_2 = 10^5 M_{\odot}$, 
(d) $m_1 = 10^6 M_{\odot}$; $m_2 = 10^6 M_{\odot}$,
(e) $m_1 = 10^6 M_{\odot}$; $m_2 = 10^5 M_{\odot}$,  
and  (f) $m_1 = 10^5 M_{\odot}$; $m_2 = 10^5 M_{\odot}$. Figure \ref{Fig.MCs}
and table \ref{Tab.MCs} summarize the results.

The key result, is that using the FWF the errors are smaller than with the RWF.
There are big improvements for the 
$10^7 M_{\odot}-10^7 M_{\odot}$, and $10^7 M_{\odot}- 10^6 M_{\odot}$ in 
angular resolution and  distance measurement:
the angular resolution improves in average
25 and 7.3, respectively; and  the luminosity distance by factors of 62 and 6.7,
respectively. One should also notice that, in those two cases, 
 those parameters were poorly determined 
using only the RWF. For the other sets of masses the averaged improvement in
angular resolution and luminosity distance are more moderate, between $1.3-3$ for $\Delta\Omega_N$,
and $1.7-2.7$ for $\Delta D_L/D_L$.
In all cases the masses are determined much more accurate, even by several orders of magnitude in the
case of $\mu$, using the FWF. For the equal mass cases, the errors in $\Mc$ and $\mu$ are of
 the same order using the FWF.

Because of the different harmonics, the FWF has a much greater richness than RWF 
 that clearly improves the parameter estimation.
It is worth mentioning that similar level of improvements were obtained in \cite{Sintes:1999ch}
where the waveform considered was only at the 0.5PN-2PN order in amplitude and phase, respectively,
i.e., adding the first and third harmonics. This suggests that the improvement in parameter estimation
is mainly due to the inclusion of the third harmonic, which 
also increases the mass reach of LISA.
The importance of the different PN orders has been discussed in 
detailed recently by Arun {\it et al.} in 
\cite{Arun:2007qv,Arun:2007p} for some particular cases, 
and  also  for ground-based
detectors \cite{Van Den Broeck:2006ar}. 

In Figure~\ref{Fig.hist76semi}
we compare the distribution of the measurement errors for 
a waveform containing only the second and third harmonics, keeping both amplitude and
phase at the 2PN order, with the FWF and the RWF, for the pair of masses
$m_1 = 10^7 M_{\odot}$ and $m_2 = 10^6 M_{\odot}$. This figure shows
how a substantial improvement in parameter estimation is obtained by adding the third harmonic.
In this case, the mean and standard deviation of the logarithm of the measurement errors
 are the following: $\log_{10} \Delta\Omega_N =  -2.33\pm 1.06$ srad, 
 $\log_{10} \Delta\Omega_L = -2.05\pm 0.46$ srad,
 $\log_{10} \Delta D_L/D_L = -1.22\pm 0.27$,
 $\log_{10} \Delta tc = 2.96\pm 0.19$ s,
 $\log_{10} \Delta \Mc/\Mc = -2.44\pm 0.17$ and
 $\log_{10} \Delta \mu/\mu = -1.57\pm 0.36$.
These results are very close to those obtained 
for the FWF as can be seen from table \ref{Tab.MCs} and are also 
in agreement with the level of
improvement found in \cite{Arun:2007p} when considering only the 0.5 PN order in amplitude.

\subsection{Pre-merger localization}

\begin{figure}
\includegraphics[width=9cm]{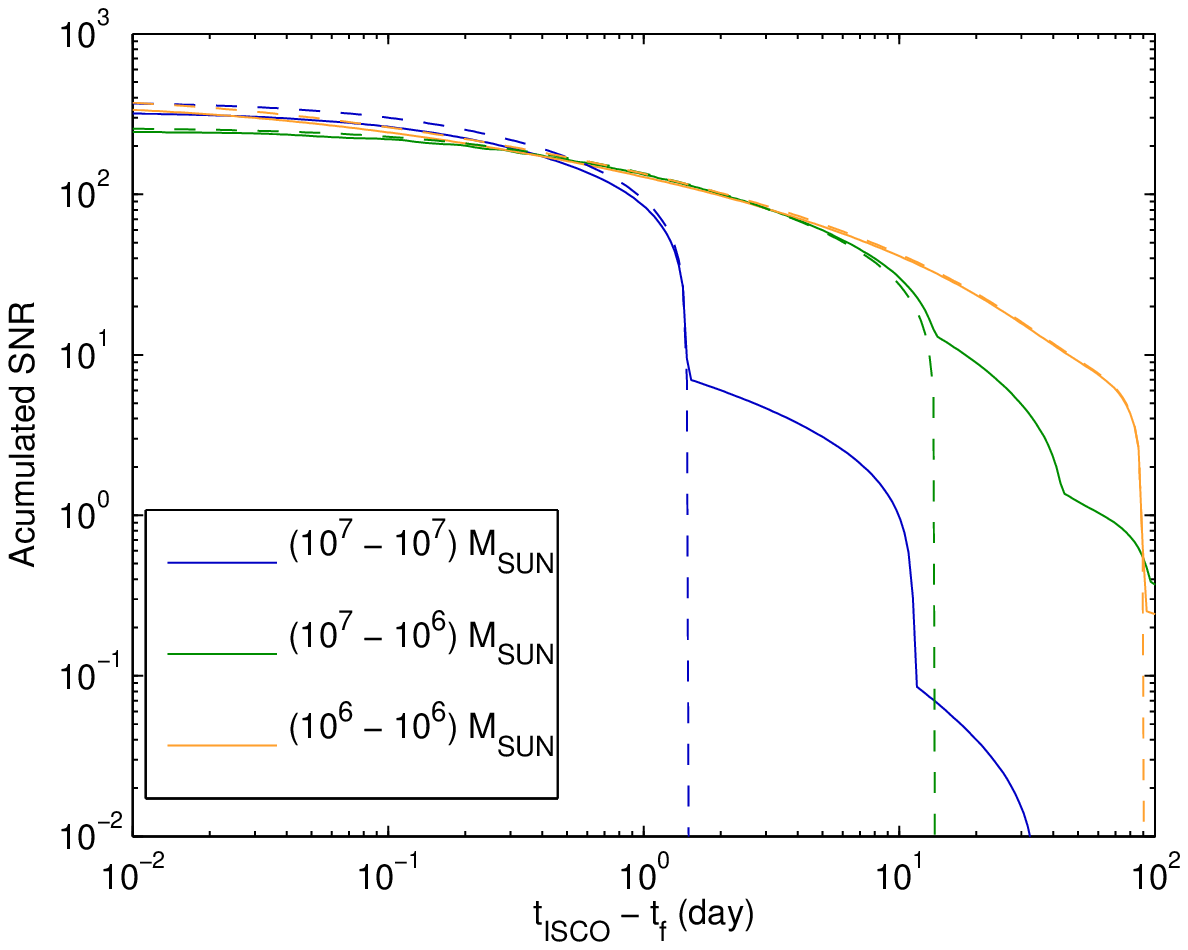} 
\caption{The progressive accumulation of SNR as a function of a look-back time;
The observations refer to the final year
of the inspiral of supermassive black hole as recorded by LISA for fiducial
sources at redshift $z=1$, with $\cos \theta_N = -0.6$, $\phi_N = 1$, $\cos \theta_L = 0.2$ and  
$\phi_L = 3$.
The solid lines correspond the FWF and dashed lines to RWF.}
\label{Fig.Sirens1}
\end{figure}

\begin{figure}
 \includegraphics[width=9cm]{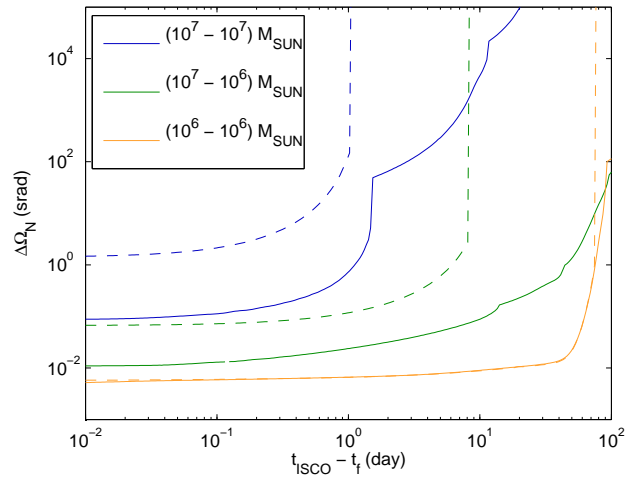} 
\caption{Time-dependence angular resolution as a function of  a look-back time
for the same sources as in Fig.~\ref{Fig.Sirens1}.
The solid lines correspond the FWF and dashed lines to RWF.}
\label{Fig.Sirens}
\end{figure}

From an astronomical point of view, one of the most attractive features is the possibility 
that LISA might have enough angular resolution to locate
the galaxy or galaxy cluster where the coalescence of a massive black hole takes
place and therefore identify potential electromagnetic counterparts.
The angular resolution is deduced primarily from the detector's motion around the Sun, so
one expects that the uncertainty in the angular resolution will not change so much
during the last days before merger.
Therefore, as discussed in \cite{Kocsis}, we are interested in analyzing the 
time-dependence of the 
angular resolution and SNR, as a function of some look-back time $t_{\rm ISCO}-t_f$ prior to coalescence, and
measure the  importance of the FWF versus RWF.

For fiducial sources at $z=1$ and a given sky location and orientation, Figures \ref{Fig.Sirens1}
and \ref{Fig.Sirens} confirm the importance of the FWF, in particular for those systems with
a higher total mass: $10^7 M_{\odot}-10^7 M_{\odot}$ and $10^7 M_{\odot}- 10^6 M_{\odot}$, 
while there
is not much difference for  the $10^6 M_{\odot}- 10^6 M_{\odot}$ one.
For  equal masses,  only the even multipoles (2, 4, 6) contribute to the FWF, while for 
 the unequal masses there are contribution from  all the six harmonics.
The 'jumps' in the progressive accumulation of SNR 
correspond to those times in which a new higher harmonic 
\textit{enters} into the LISA band and it is 
related to the lower frequency cut-off at $5\times 10^{-5}$~Hz we have imposed.
For example, in the $10^7 M_{\odot}-10^7 M_{\odot}$ case the contribution 
of the 4th harmonic becomes relevant
around 10 days before coalescence while the 2nd harmonic rapidly increases the SNR 2 days before
coalescence. For the unequal mass case $10^7 M_{\odot}- 10^6 M_{\odot}$ we clearly see the
contributions of the 2nd, 3rd and 4th harmonics.

Using the FWF, not only the uncertainty in sky location decreases but also  allows
earlier warnings.


\section{Summary and outlook}
\label{sec:summary}


We have considered LISA observations of supermassive black hole systems in the final stage of
inspiral. We have restricted our analysis to systems in circular orbit with negligible spins, modeling
the radiation  at the full 2PN order, and we have compared with the restricted-2PN.
With both waveform models we have determined the mean-square errors associated with the 
parameter measurements of black hole binaries in the mass range  $10^8\Ms-10^5\Ms$, 
for equal and unequal mass cases, for a wide range of source locations and orientations.

The conclusions of this work are particularly important with regard to the astrophysical reach of
future LISA measurements. Our analysis clearly shows that modeling the inspiral with the full
post-Newtonian waveforms, as compared to the restricted-PN ones, not only extends the reach to 
higher mass systems  up to $10^8\Ms$,
as previously shown in \cite{Arun:2007qv}, but also improves in general the 
 parameter estimation, and allows for early warnings for  systems with  a
high total mass. There are remarkable improvements in angular resolution and distance 
measurement for systems with a total mass higher than $5\times10^6\Ms$, as well as a large 
improvement in the mass determination. For  $\Delta\mu/\mu$, the improvement is more than three orders
of magnitude in the case of equal masses.

For binary systems of $10^7 M_{\odot}-10^7 M_{\odot}$, 
and $10^7 M_{\odot}- 10^6 M_{\odot}$ at redshift $z=1$,
the angular resolution improves in average
25 and 7.3, respectively; and  the luminosity distance by factors of 62 and 6.7,
respectively. Moreover,  for the equal mass case $10^7 M_{\odot}-10^7 M_{\odot}$,
for all those sky locations in which
we obtained the best angular resolution,  $\Delta\Omega_N\sim 10^{-4}$~srad with the FWF, the 
gain in $\Delta\Omega_N$ is up to 3 orders of magnitude. These results are in agreement with those
recently found in \cite{Arun:2007p}.

These improvements are related to the fact that the FWF has a much greater richness than the RWF,
due to the presence of the higher harmonics and, in particular, the main contribution
to this improvement is associated to the third harmonic.

There are a number of issues that influence these observations: the instrumental lower-frequency
cut-off we have imposed and the confusion noise model we have used are two of them.
It would also be very interesting to revise these conclusions, by considering black holes
with large spins and precession, since it is known that the presence of spins reduces the errors
with which the source parameters are measured \cite{Vecchio:2003tn}.

Another issue is the fact that the largest improvement happens for systems
with a total mass of  $5\times10^6\Ms$ and higher, and although, we are still in a
 regime of large SNR, one could question the validity of the Fisher matrix approach. 
 Other investigations are currently underway using alternative methods \cite{bps}.

\begin{appendix}

\section{Mass parameters transformation}
\label{sec:equalmass}

\begin{figure}
\includegraphics[width=7cm]{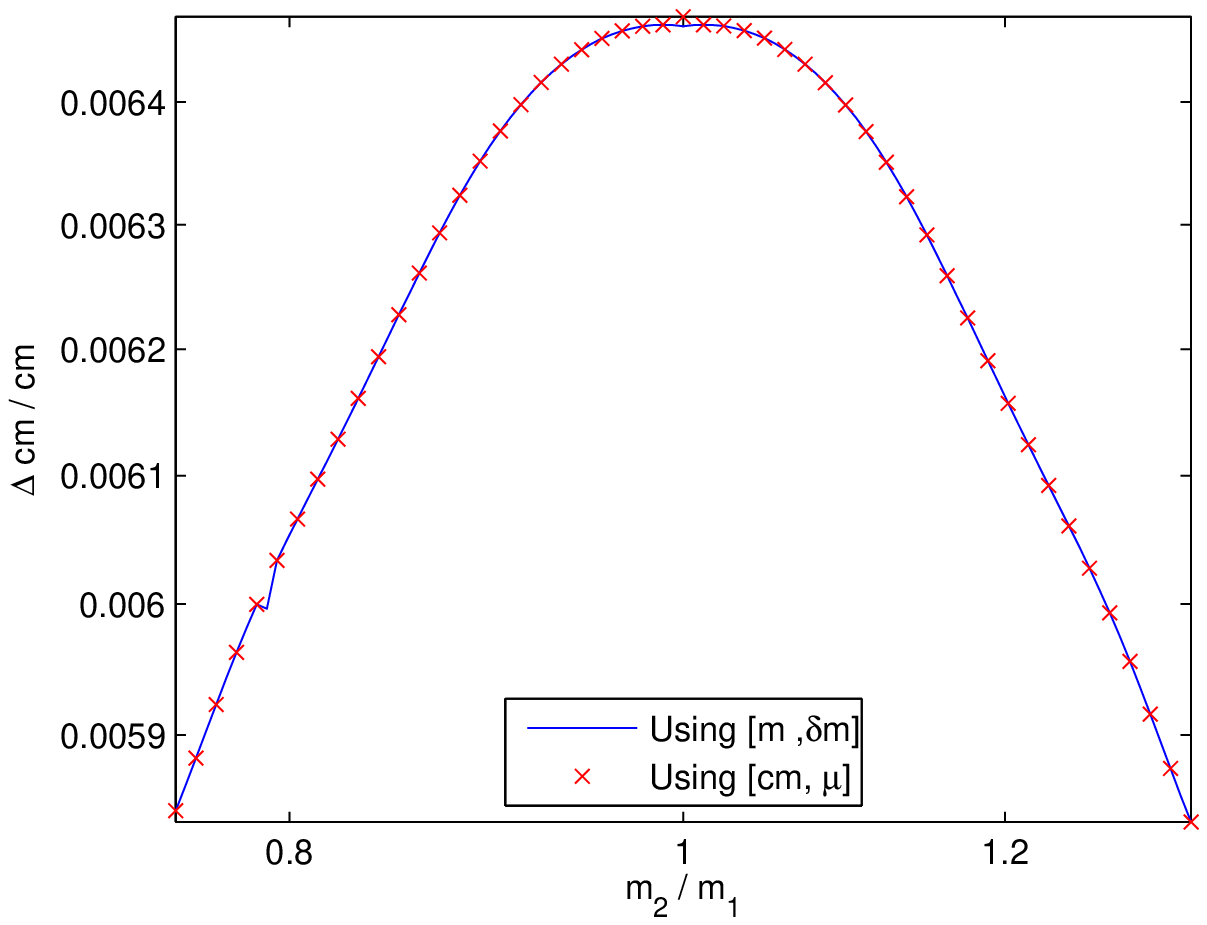} \\
 \includegraphics[width=7cm]{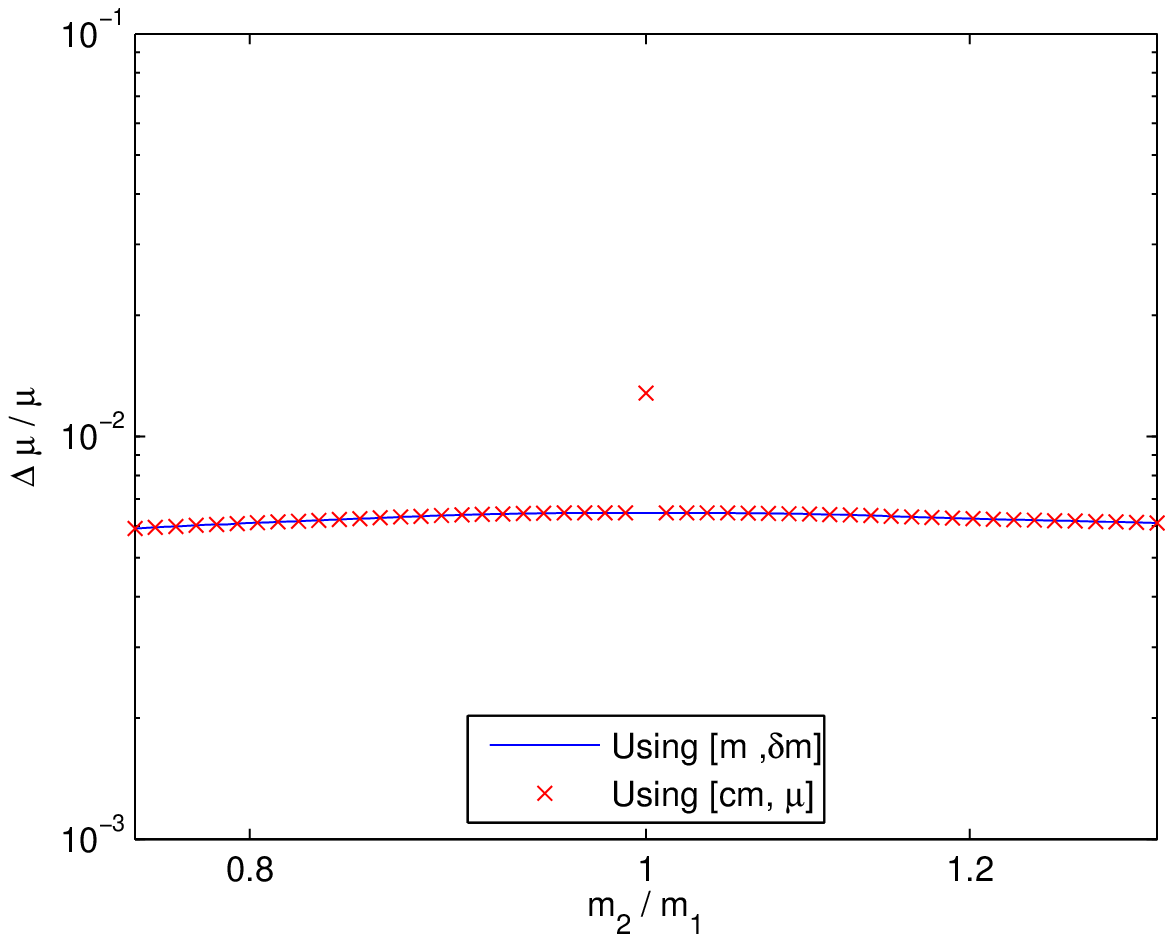}
\caption{Mass measurements using the  FWF  for binary systems with a total mass $M= 10^7 M_{\odot}$.
The solid lines correspond to compute the errors in $\{ \delta m, M\}$ and converting them,
by mean of equations (\ref{6.2})-(\ref{6.8}),  into $\{\Mc,\mu\}$ , and crosses correspond to compute
the errors in $\{\Mc,\mu\}$  directly.
 }
\label{Fig.ChngVarFWF}
\end{figure}

\begin{figure}
\includegraphics[width=7cm]{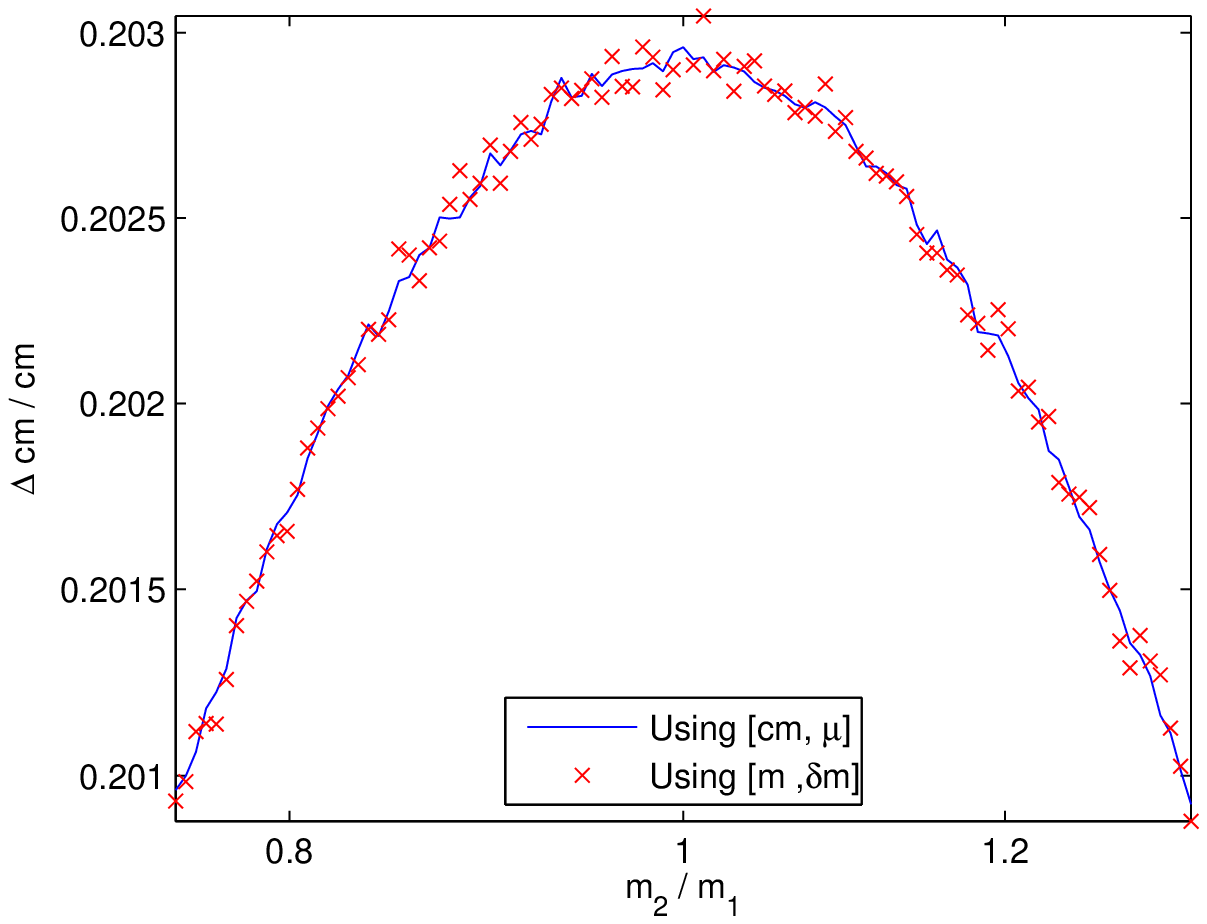} \\
\includegraphics[width=7cm]{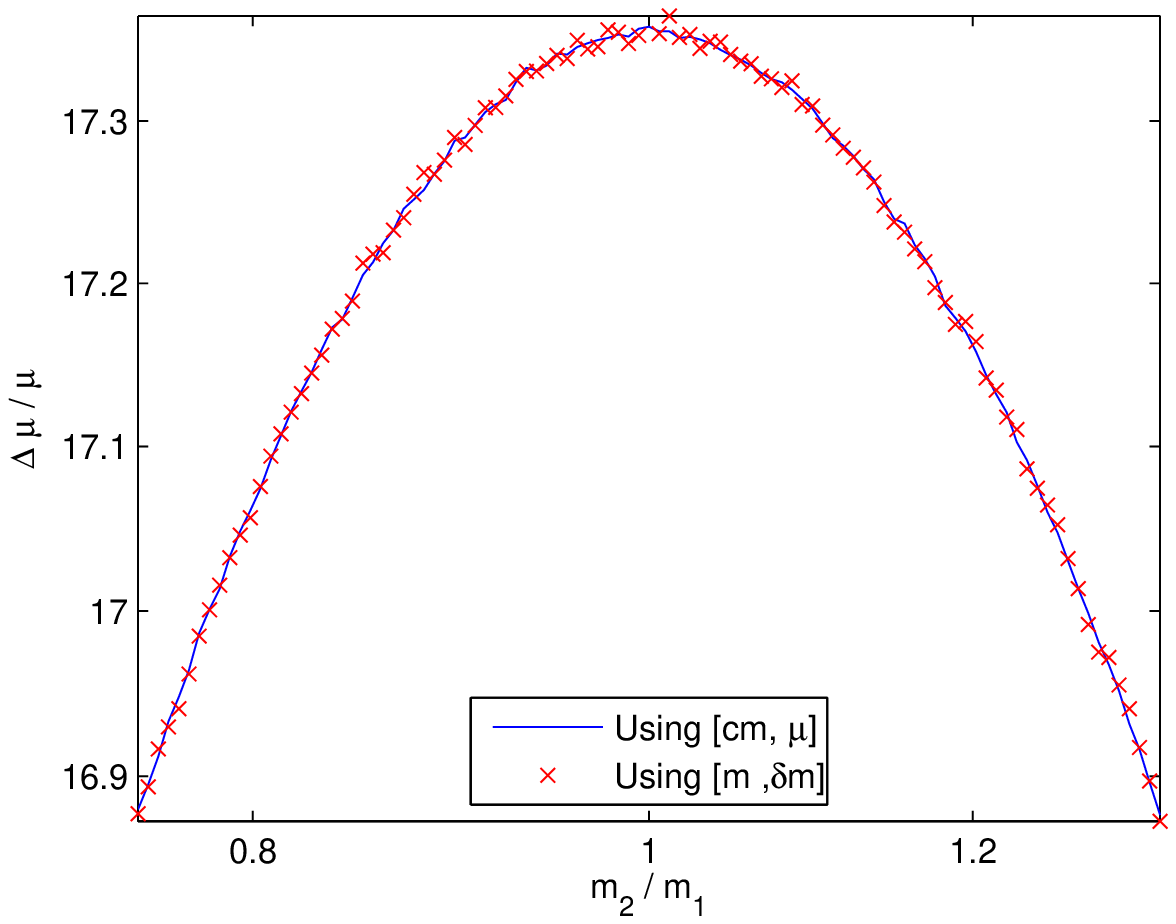}
\caption{
Mass measurements using the  RWF  for binary systems with a total mass $M= 10^7 M_{\odot}$.
The solid lines correspond to compute the errors in $\{\Mc,\mu\}$  directly, and 
crosses correspond to compute
the errors in $\{ \delta m, M\}$ first and converting them into $\{\Mc,\mu\}$.
}
\label{Fig.ChngVarRWF}
\end{figure}

\begin{figure}
\begin{center}
\includegraphics[width=9cm]{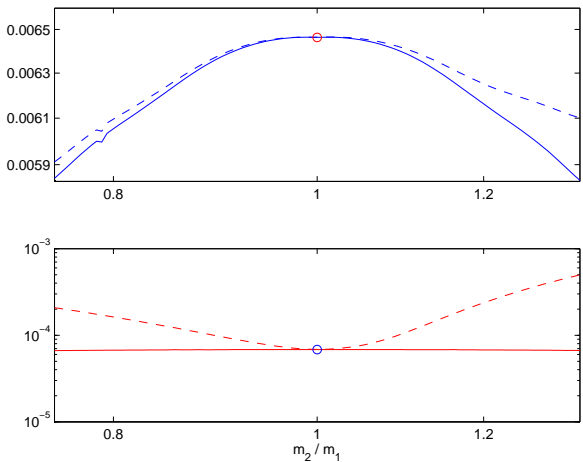}
\end{center}
\caption{Study of the mass parameter measurements in the limit  $\delta m \to 0$ using the FWF
with the $\{ \delta m, M\}$ mass parameterization for systems with a total 
mass  (top)  $M = 10^7 M_{\odot}$ and (bottom) $M = 10^6 M_{\odot}$.
In both panels the solid lines correspond to $\Delta~\cm/\cm$, the dashed lines to 
$\Delta~\mu/\mu$ and the circle indicates the equal mass case.}
\label{Fig.Limdm0}
\end{figure}

As discussed earlier in Sec.~\ref{sec:results} it is traditional to parameterize the masses using
$\ln\Mc$ and $\ln\mu$ because of their appearance in the waveform phase. However the higher order PN
amplitude terms depend on the mass difference $\delta m$ and total mass  $M$. It is a matter of
choice to work with one or another set of parameters. Errors in  $\{\Mc,\mu\}$ can be computed easily,
in principle,
given the errors in $\{ \delta m, M\}$ by mean of equations (\ref{6.2})-(\ref{6.8}). Unfortunately
the Jacobian of the transformation between $\{\Mc,\mu\}$  and $\{ \delta m, M\}$ is singular when 
$ \delta m=0$ leading to problems in evaluating the Fisher matrix.

For  unequal masses, we find that computing the
errors in  $\{ \delta m, M\}$ and then converting gives the same result as simply computing the
errors in  $\{\Mc,\mu\}$ directly, and this result is independent of the waveforms used FWF or RWF,
as it is shown in figures \ref{Fig.ChngVarFWF} and \ref{Fig.ChngVarRWF}. However, depending of our
choice of mass parameters and waveform model we use,  
 there appear divergences when evaluating
the Fisher matrix for equal masses. In particular, for the RWF
we do not  trust the $\{ \delta m, M\}$ parameterization, and for the FWF we do not trust the 
$\{\Mc,\mu\}$. That is, we trust the solid lines in Fig.~\ref{Fig.ChngVarFWF} and 
\ref{Fig.ChngVarRWF}, but not the crosses for equal masses, that either do not coincide 
with the solid lines for the FWF or could not even be computed using the RWF.
Because the solid lines did not present any misbehavior nor discontinuities when approaching the
equal mass case, we decided to use (independently of the masses) the $\{\Mc,\mu\}$ parameterization
for the RWF and the $\{ \delta m, M\}$ for the FWF.

Another aspect we want to study in more detail, is the fact that using the FWF,
 for equal masses, 
$\Delta~\cm/\cm$ and $\Delta~\mu/\mu$ become identically the same, but this is not true for the
RWF, nor for unequal masses. This is a consequence of setting 
$ \delta m=0$ in Eqs.~(\ref{6.2})-(\ref{6.8}). In Figure~\ref{Fig.Limdm0}
we show that this is not an artifact of
our mass transformation due to the Jacobian being singularity  at that point
but a fact.


\section{Some expressions related with the 2PN inspiral waveform}
\subsection{Time, frequency and phase evolution}
\label{Sec.tfp_evolution}

A coalescing binary system evolves by loosing energy and angular momentum ${\textbf L}$
through emission of gravitational waves of increasing frequency and amplitude. 
Working with the post-Newtonian approximation, 
taking into account possible spin motions of each object, ${\textbf S_1}$ and ${\textbf
S_2}$,
the signal frequency of the
second harmonic, $F=2 f_{orb}$, evolves, up to 2PN order,
according to \cite{Berti:2004bd,Vecchio:2003tn,Lang:2006bz,
Van Den Broeck:2006qu,Van Den Broeck:2006ar,Blanchet:1996pi}
\begin{eqnarray}\label{Eq.dF_dt}
\dfrac{dF}{dt} & = & \dfrac{96}{5} \pi F^2 \eta x^5 \left[ 1-\left(
\frac{743}{336}+\frac{11}{4} \eta\right) x^2 + \left( 4\pi-\beta\right) x^3 \right.
\nonumber \\
 & & \left. + \left( \frac{34103}{18144} + \frac{13661}{2016}\eta + \frac{59}{18}\eta^2
+\sigma\right) x^4\right] \, ,
\end{eqnarray}
where $x$ is the PN expansion parameter defined in Eq.~(\ref{Eq.xdef}),
%
and
 $\beta$ and $\sigma$ are the so called \emph{spin-orbit} and \emph{spin-spin}
parameters, respectively
\begin{equation}\label{Eq.betadef}
\beta = \dfrac{1}{12} \sum_{i=1}^{2}\left[ 113\left( \frac{m_i}{M}\right)^2 + 75\eta\right]
\left( \hat{{\textbf L}} \cdot \dfrac{{\textbf S_i}}{m_i^2}\right) \, ,
\end{equation}
\begin{equation}\label{Eq.sigmadef}
\sigma = \dfrac{\eta}{48} \left[ -247\left( \dfrac{{\textbf S_1}}{m_1^2} \cdot
\dfrac{{\textbf S_2}}{m_2^2}\right) + 721 \left( \hat{{\textbf L}} \cdot \dfrac{{\textbf
S_1}}{m_1^2}\right)~\left( \hat{{\textbf L}} \cdot \dfrac{{\textbf
S_2}}{m_2^2}\right)\right] \, .
\end{equation}
Integrating Eq.~(\ref{Eq.dF_dt}), one can derive the time evolution of the gravitational
radiation
\begin{eqnarray}\label{Eq.t(F)}
t(F) & = & t_c - \dfrac{5}{256} \dfrac{1}{\eta \pi F} x^{-5} \left[ 1+\frac{4}{3} \left(
\frac{743}{336}+\frac{11}{4} \eta\right) x^2 \right. \nonumber \\
 & & \left. - \frac{8}{5} \left( 4\pi - \beta\right) x^3 \right. \\
& & \left. + 2 \left( \frac{3058673}{1016064} + \frac{5429}{1008} \eta + \frac{617}{144}
\eta^2 - \sigma\right) x^4\right] \, , \nonumber
\end{eqnarray}
 and the phase evolution of the gravitational waveform. For the second harmonic this is
\begin{eqnarray}\label{Eq.Phi(F)}
\Phi & = & \Phi_c - \dfrac{3}{128} \dfrac{1}{\eta} x^{-5} \left[ 1 + \frac{20}{9} \left(
\frac{743}{336} + \frac{11}{4} \eta \right) x^2 \right. \nonumber \\
 & & - \left. \left( 16\pi-4\beta\right) x^3 \right.  \\
 & & + \left. 10 \left( \frac{3058673}{1016064} + \frac{5429}{1008} \eta + \frac{617}{144}
\eta^2 - \sigma\right) x^4\right] \, . \nonumber
\end{eqnarray}

The gravitational waveform can be computed in the frequency domain 
using the stationary phase approximation
\begin{equation}\label{Eq.SPA2}
\tilde{h}(\nu)= \sum_{j=1}^6 \left[ \dfrac{\tilde{h}_{j}}{2} 
e^{i\left[ \frac{j}{2}(2\pi F t_c-\Phi)-\pi/4-\varphi_{p,j}-\varphi_D \right]}
\sqrt{\dfrac{2}{j}\dfrac{1}{\frac{dF}{dt}}}\right]_{F=\frac{2}{j}\nu}
\end{equation}
where $\tilde{h}_j \equiv \dfrac{\sqrt{3}}{2} 2 M\eta \dfrac{1}{D_L}x^2 A_j$.
%
Using  Eq.~(\ref{Eq.dF_dt}),  $\left(dF/dt\right)^{-1/2}$ can be written to 2PN order as
\begin{equation}\label{Eq.1/sqrt(dF_dt)}
\dfrac{1}{\sqrt{\frac{dF}{dt}}} = \sqrt{\dfrac{5\pi}{96}} M \eta^{-1/2} x^{-11/2} \left(
\sum_{n=0}^4 k_n x^n\right) \, ,
\end{equation}
where coefficients $k_n$ are defined as follows
\begin{eqnarray}
k_0 &=& 1 \nonumber\\
k_1 &=& 0 \nonumber\\
k_2 &=& \dfrac{1}{2} \left( \frac{743}{336}+\frac{11}{4}\eta\right) \nonumber\\
k_3 &=& -\dfrac{1}{2} \left( 4\pi - \beta\right) \nonumber\\
k_4 &=& \frac{7266251}{8128512} + \frac{18913}{16128} \eta + \frac{1379}{1152}\eta^2 -
\frac{\sigma}{2} \, .
\end{eqnarray}

\subsection{$\hat{u}_{(+,\cross),j}$ and $\hat{w}_{(+,\cross),j}$ up to 2PN}
\label{Sec.us_ws}

In Sec.~\ref{sec:pn} we have seen a general form to expand the GW 
amplitude, of a particular
multipole, as a summation of its different PN contributions. The
analytical expression of all the terms appearing in  Eq.~(\ref{Eq.us_ws})
can be obtained from \cite{Blanchet:1996pi}.
Here we explicitely give all the non-vanishing terms 
$\hat{u}_{(+,\cross),j}^{(n)}$ and $\hat{w}_{(+,\cross),j}^{(n)}$ 
up to $n=4$.

%

\begin{center}
\emph{Contributions to $\hat{u}_{+,j}$}
\end{center}
\begin{eqnarray*}
\hat{u}_{+,1}^{(1)} & = & -\dfrac{1}{8} (5+c^2) \\
\hat{u}_{+,1}^{(3)} & = & \dfrac{1}{192} \left[(57+60c^2-c^4)-2\eta(49-12c^2-c^4)\right] \\
\hat{u}_{+,1}^{(4)} & = & -\dfrac{\pi}{8} (5+c^2) \\
\end{eqnarray*}
\begin{eqnarray*}
\hat{u}_{+,2}^{(0)} & = & -(1+c^2) \\
\hat{u}_{+,2}^{(2)} & = & \dfrac{1}{6} \left[(19+9c^2-2c^4)-\eta(19-11c^2-6c^4)\right] \\
\hat{u}_{+,2}^{(3)} & = & -2\pi (1+c^2) \\
\hat{u}_{+,2}^{(4)} & = & \dfrac{1}{120} \left[ (22+396c^2+145c^4-5c^6) \right. \\
 & & \left. + \frac{5}{3}\eta(706-216c^2-251c^4+15c^6) \right. \\
 & & \left. - 5\eta^2(98-108c^2+7c^4+5c^6)\right]
\end{eqnarray*}
\begin{eqnarray*}
\hat{u}_{+,3}^{(1)} & = & \dfrac{9}{8} (1+c^2) \\
\hat{u}_{+,3}^{(3)} & = & -\dfrac{9}{128} \left[(73+40c^2-9c^4)-2\eta(25-8c^2-9c^4)\right]
\\
\hat{u}_{+,3}^{(4)} & = & \dfrac{27}{8} \pi(1+c^2) 
\end{eqnarray*}
\begin{eqnarray*}
\hat{u}_{+,4}^{(2)} & = & -\dfrac{4}{3} (1+c^2) (1-3\eta) \\
\hat{u}_{+,4}^{(4)} & = & \dfrac{2}{15} \left[(59+35c^2-8c^4) \right. \\
 & & \left. - \frac{5}{3}\eta(131+59c^2-24c^4) \right. \\
 & & \left. + 5\eta^2(21-3c^2-8c^4) \right] 
\end{eqnarray*}
\begin{eqnarray*} 
\hat{u}_{+,5}^{(3)} & = & \dfrac{625}{384} (1+c^2) (1-2\eta) \\ \\
\hat{u}_{+,6}^{(4)} & = & -\dfrac{81}{40} (1+c^2) (1-5\eta+5\eta^2)
\end{eqnarray*}
\begin{center}
\emph{Contributions to $\hat{w}_{+,j}$}
\end{center}
\begin{eqnarray*}
\hat{w}_{+,1}^{(4)} & = & \dfrac{1}{40} \left[11+7c^2+10(5+c^2)\ln 2\right] \\
\\
\hat{w}_{+,3}^{(4)} & = & -\dfrac{27}{40} (1+c^2) \left[7-10\ln \frac{3}{2}\right]
\end{eqnarray*}
\begin{center}
\emph{Contributions to $\hat{w}_{\cross,j}$}
\end{center}
\begin{eqnarray*}
\hat{w}_{\cross,1}^{(1)} & = & -\dfrac{3}{4} c \\
\hat{w}_{\cross,1}^{(3)} & = & \dfrac{c}{96} \left[(63-5c^2) - 2\eta(23-5c^2)\right] \\
\hat{w}_{\cross,1}^{(4)} & = & -\dfrac{3\pi}{4} c
\end{eqnarray*}
\begin{eqnarray*}
\hat{w}_{\cross,2}^{(0)} & = & -2c \\
\hat{w}_{\cross,2}^{(2)} & = & \dfrac{c}{3} \left[(17-4c^2) - \eta(13-12c^2)\right] \\
\hat{w}_{\cross,2}^{(3)} & = & -4\pi c \\
\hat{w}_{\cross,2}^{(4)} & = & \dfrac{c}{60} \left[(68+226c^2-15c^4) \right. \\
 & & \left. + \frac{5}{3}\eta(572-490c^2+45c^4) \right. \\
 & & \left. - 5\eta^2(56-70c^2+15c^4)\right]
 \end{eqnarray*}
\begin{eqnarray*}
\hat{w}_{\cross,3}^{(1)} & = & \dfrac{9}{4}c \\
\hat{w}_{\cross,3}^{(3)} & = & -\dfrac{9}{64}c \left[(67-15c^2) - 2\eta(19-15c^2) \right] \\
\hat{w}_{\cross,3}^{(4)} & = & \dfrac{27}{4}\pi c 
\end{eqnarray*}
\begin{eqnarray*}
\hat{w}_{\cross,4}^{(2)} & = & -\dfrac{8}{3} c\, (1-3\eta) \\
\hat{w}_{\cross,4}^{(4)} & = & \dfrac{4}{15} c \left[(55-12c^2) -
\frac{5}{3}\eta(119-36c^2) \right. \\
 & & \left. + 5\eta^2(17-12c^2) \right] 
 \end{eqnarray*}
\begin{eqnarray*}
\hat{w}_{\cross,5}^{(3)} & = & \dfrac{625}{192} c\, (1-2\eta) \\
\\
\hat{w}_{\cross,6}^{(4)} & = & -\dfrac{81}{20} c\, (1-5\eta+5\eta^2)
\end{eqnarray*}
\begin{center}
\emph{Contributions to $\hat{u}_{\cross,j}$}
\end{center}
\begin{eqnarray*}
\hat{u}_{\cross,1}^{(4)} & = & -\dfrac{3}{20} c\, (3+10\ln 2) \\
\\
\hat{u}_{\cross,3}^{(4)} & = & \dfrac{27}{20} c\, (7-10\ln \frac{3}{2})
\end{eqnarray*}

\end{appendix}

\section*{Acknowledgments}

We are especially grateful to Alberto Vecchio for encouragement, 
useful discussions
and for graciously sharing with us his LISA parameter estimation code.
We also thank B.S. Sathyaprakash, Bernard Schutz,  K. G. Arun, Stanislav Babak,
Chris van den Broeck, Daniel Holz, Scott  Hughes,  Bala Iyer and  Sascha Husa,
for many useful discussions.
This work was supported by the 
Spanish Ministerio de Educaci\'on y Ciencia research project
FPA-2004-03666, and the 'Conselleria d'Economia Hisenda i Innovaci\'o' of the
Government of the Balearic Islands.
AMS acknowledges the Albert Einstein Institute and the University of
Jena for hospitality during several stages of this work.

\end{document}